\documentclass[reqno]{amsart}
\usepackage{times}

\usepackage{amsmath}
\usepackage{amssymb}
\usepackage{amsfonts}
\usepackage{latexsym}
\usepackage{textcomp}
\usepackage{verbatim}

\newtheorem{lemma}{Lemma}

\newtheorem{theorem}{Theorem}

\newcommand\abs[1]{| #1 |}
\newcommand\Abs[1]{\left| #1 \right|}
\newcommand\norm[1]{\| #1 \|}
\newcommand\Norm[1]{\left\| #1 \right\|}
\newcommand\trnorm[1]{\| #1 \|_\intercal}
\newcommand\cbnorm[1]{\| #1 \|_{\rm cb}}
\newcommand\ket[1]{| #1 \rangle}
\newcommand\bra[1]{\langle #1 |}
\newcommand\braket[2]{\langle #1 | #2 \rangle}
\newcommand\ketbra[2]{| #1 \rangle \langle #2 |}
\newcommand\Bigbar[2]{\mathrel{\left|\vphantom{#1}#2\right|}}
\newcommand\Braket[3]{\left\langle #1 \Bigbar{#1}{#2} #3\right\rangle}
\newcommand\map[3]{#1 : #2 \longrightarrow #3}
\newcommand\tran[1]{\widetilde{#1}}
\def\co{\operatorname{co}}
\def\ran{\operatorname{ran}}
\renewcommand\Re{\operatorname{Re}}
\renewcommand\Im{\operatorname{Im}}
\def\Sp{\operatorname{Sp}}
\def\Tr{\operatorname{Tr}}
\def\id{{\rm id}}
\def\opt{\mathrm{o}}
\def\openone{\hbox{\upshape \small1\kern-3.3pt\normalsize1}}
\def\idty{\openone}
\def\tp{\otimes}
\def\cA{{\mathcal A}}
\def\cB{{\mathcal B}}
\def\cC{{\mathcal C}}
\def\cD{{\mathcal D}}

\def\cF{{\mathcal F}}
\def\cH{{\mathcal H}}
\def\cK{{\mathcal K}}

\def\cM{{\mathcal M}}

\def\cS{{\mathcal S}}

\def\fk{{\mathfrak k}}
\def\fh{{\mathfrak h}}
\def\fg{{\mathfrak g}}

\def\C{{\mathbb C}}
\def\N{{\mathbb N}}
\def\R{{\mathbb R}}
\begin{document}

\title{Operational Distance and Fidelity for Quantum Channels}
\author[Viacheslav P. Belavkin, Giacomo Mauro D'Ariano, Maxim
  Raginsky]
       {Viacheslav P.~Belavkin${}^{1,*}$,
         Giacomo Mauro D'Ariano${}^{2,3,\dag}$,
         Maxim Raginsky${}^{3,\ddag}$}

\address{${}^1$Department of Mathematics, University of Nottingham,
  NG7 2RD Nottingham, UK\vskip 1em
${}^2$QuIT Group, INFM Unit\`a di Pavia, Universit\`a di Pavia,
  Dipartimento di Fisica ``A. Volta'', via Bassi 6, I-27100 Pavia,
  Italy\vskip 1em
${}^3$Center for Photonic Communication and Computing, Department of
  Electrical and Computer Engineering, Northwestern University,
  Evanston, IL 60208-3118, USA\vskip 1em
${}^*$ E-mail address: {\tt vpb@maths.nott.ac.uk}\vskip 1em
${}^\dag$ E-mail address: {\tt dariano@unipv.it}\vskip 1em
${}^\ddag$ E-mail address: {\tt maxim@ece.northwestern.edu}}

\begin{abstract}
We define and study a fidelity criterion for
quantum channels, which we term the {\em minimax fidelity}, through a
noncommutative generalization of maximal Hellinger distance between
two positive kernels in classical probability theory. Like other known
fidelities for quantum channels, the minimax
fidelity is well-defined for channels between
finite-dimensional algebras, but it also applies to a certain class of
channels between infinite-dimensional algebras (explicitly, those
channels that possess an operator-valued Radon--Nikodym density with
respect to the trace in the sense of Belavkin--Staszewski) and induces
a metric on the set of quantum channels which is topologically
equivalent to the CB-norm distance between channels, precisely in the
same way as the Bures metric on the density operators associated with
statistical states of quantum-mechanical systems, derived from the
well-known fidelity (`generalized transition probability') of Uhlmann,
is topologically equivalent to the trace-norm distance.\\


\end{abstract}
\maketitle
\tableofcontents

\thispagestyle{empty}

\section{Introduction}
\label{sec:intro}

Many problems in quantum information science \cite{Key02,NC00}, both in
theory and in experiment, involve finding a set of quantum-mechanical
states or channels that solve some sort of an optimization problem,
typically formulated in terms of a numerical criterion that measures
how close a given pair of states or operations are to each other. (Many such
criteria have been proposed to date, each defined with
specific theoretical or experimental considerations in mind; see
Ref.~\cite{GLN04} for a recent comprehensive survey.)

Let us first consider the case of quantum states, i.e., density
operators. Let $\fh$ be a complex
separable Hilbert space associated to a quantum-mechanical
system. Given a pair of density operators $\rho,\sigma$, i.e.,
positive trace-class operators with unit trace, one can use either the
\emph{fidelity} \cite{Alb03,AU00,Joz94,Uhl76}
\begin{equation}
F(\rho,\sigma):=\Tr\big[(\rho^{1/2}\sigma\rho^{1/2})^{1/2}\big]
\label{eq:statefid}
\end{equation}
or the \emph{trace-norm (half-) distance}
\begin{equation}
D(\rho,\sigma):=\frac{1}{2}\trnorm{\rho-\sigma},
\label{eq:tnd}
\end{equation}
where $\trnorm{\rho} := \Tr \abs{\rho}$, $\abs{\rho} := (\rho^\dag
\rho)^{1/2}$ \cite{Sch60,Sim79}. Loosely speaking, two states $\rho$ and $\sigma$ are
close to each other if $F(\rho,\sigma)$ is large, or if $D(\rho,\sigma)$ is
small. In fact, as follows from the key inequality \cite{AU00,FG99}
\begin{equation}
1-F(\rho,\sigma)\leq D(\rho,\sigma)\leq\sqrt{1-F^2(\rho,\sigma)},
\label{eq:fg}
\end{equation}
the fidelity and the trace-norm distance are equivalent in the sense that any
two density operators that are close to one another in the sense of
(\ref{eq:statefid}) are also close in the sense of (\ref{eq:tnd}), and vice versa.

As for \emph{quantum channels}, i.e., normal completely positive unital
mappings from an operator
algebra $\cB={\cB}(\fh)$ into another algebra $\cA=\cB\left(  \fg\right)  $, where
$\fg$ and {$\fh$} are complex separable Hilbert spaces,
things get somewhat complicated. Consider, for instance, the case when
$\fg$ is finite-dimensional, and let $m := \dim\fg$. Fix an orthonormal basis $\{|j\rangle\}_{j=1}^{m}$ of
$\fg$, and let $|\psi\rangle:=m^{-1/2}\sum_{j=1}^{m}|j\rangle
\otimes|j\rangle$ be the normalized maximally entangled vector in the
product space $\fg \otimes\fg$. Given two quantum channels
$\map{\Phi,\Psi}{\cB}{\cA}$, one can measure their closeness in terms
of the fidelity of the states on $\cB\otimes\cA$, obtained from the maximally entangled
state $\pi=|\psi\rangle\langle\psi|$ by applying the predual channels $\Phi_{\intercal}$ and
$\Psi_{\intercal}$ (cf.~Section~\ref{sec:prelims} for precise
definitions) to the first factor in the tensor product:
\begin{eqnarray*}
&& \Phi_{\intercal}\otimes\mathrm{id} (  \pi)
= \frac{1}{m}\sum_{i=1}^{m}\sum_{k=1}^{m}\Phi_{\intercal}\left(  |i\rangle
\langle k|\right)  \otimes|i\rangle\langle k|\equiv\rho\\
&& \Psi_{\intercal}\otimes\mathrm{id} (  \pi)
= \frac{1}{m}\sum_{i=1}^{m}\sum_{k=1}^{m}\Psi_{\intercal}\left(  |i\rangle
\langle k|\right)  \otimes|i\rangle\langle k|\equiv\sigma.
\end{eqnarray*}
The fidelity $F\left(  \rho,\sigma\right)  $, taken as the {\em channel fidelity}
\begin{equation}
{\mathcal{F}}(\Phi,\Psi):=F\big(\Phi_{\intercal}\otimes\mathrm{id}(
\pi)  ,\Psi_{\intercal}\otimes\mathrm{id}(  \pi)
\big)
\label{eq:chfid}%
\end{equation}
by Raginsky in Ref.~\cite{Rag01}, enjoys many properties parallel to those of the fidelity
(\ref{eq:statefid}) for quantum states. Alternatively, one can adopt the
(half-) distance \cite{Key02,AKN97,Kit97}
\begin{equation}
{\mathcal{D}}(\Phi,\Psi):=\frac{1}{2}\|\Phi-\Psi\|_{\mathrm{cb}%
},\label{eq:cbd}%
\end{equation}
where $\cbnorm{\bullet}$ denotes the so-called {\em norm of complete
boundedness} (or CB-norm for short; cf.~Section~\ref{ssec:cbmaps} for
details). We note that the CB-norm half-distance (\ref{eq:cbd}) can be given in
terms of the
trace-norm distance (\ref{eq:tnd}) between density operators by means of
the variational expression \cite{Key02,AKN97,Kit97}
\begin{equation}
{\mathcal{D}}(\Phi,\Psi)=\sup_\pi D\big(\Phi_{\intercal}\otimes\mathrm{id}(
\pi)  ,\Psi_{\intercal}\otimes\mathrm{id}(  \pi)
\big),\label{eq:cboper}%
\end{equation}
where the supremum is taken over all density operators $\pi$ on the
tensor product space $\fg \tp \fg$. By analogy with density operators of the states, we are tempted to say that
two quantum channels, $\Phi$ and $\Psi$, are close either if ${\mathcal{F}%
}(\Phi,\Psi)$ is large or if ${\mathcal{D}}(\Phi,\Psi)$ is
small. However, in addition to the finite-dimension
restriction $\dim\fg<\infty$ [the only case under which the definition
(\ref{eq:chfid}) of the channel fidelity makes sense], we encounter the following difficulty. It turns out \cite{Rag01} that, as a
criterion of closeness, the CB-norm distance (\ref{eq:cbd}) is strictly
stronger than the fidelity measure (\ref{eq:chfid}) in the sense that even
when ${\mathcal{D}}(\Phi,\Psi)$ is large, ${\mathcal{F}}(\Phi,\Psi)$ may be
quite large as well, and may even become equal to one in the limit $\dim
\fg\longrightarrow\infty$. Consider, for instance, the case $\Psi
=\mathrm{id}$. Then one can show \cite{Rag01} that
\begin{equation}
1-\cD(\Phi,\id) \leq \cF(\Phi,\id) \leq
 \sqrt{1-(1/4)\cD^2(\Phi,\id)},
\label{eq:badbound}
\end{equation}
and we immediately see that when $\Phi$ is such that
$\mathcal{D}(\Phi,\mathrm{id})$ attains its maximum value of unity, the fidelity
${\mathcal{F}}(\Phi,\mathrm{id})$ is still bounded between $0$ and $\sqrt{3}/2$. To
make matters worse, the only bound on (\ref{eq:cbd}) in terms of
(\ref{eq:chfid}) known so far is
\begin{equation}
1\geq{\mathcal{D}}(\Phi,\Psi)\geq1-{\mathcal{F}}(\Phi,\Psi),
\label{eq:lobound}%
\end{equation}
as follows readily from Eqs.~(\ref{eq:fg}) and
  (\ref{eq:cboper}). Furthermore, one can easily find sequences $\{
  \Phi_m\}$, $\{\Psi_m\}$ of
  channels $\map{\Phi_m,\Psi_m}{\cB(\C^m)}{\cB(\C^m)}$, such that
  $\cD(\Phi_m,\Psi_m) \neq 0$ for all $m$, while
$$
\lim_{m \to \infty}{\mathcal{F}}(\Phi_m,\Psi_m)=1.
$$
Indeed, consider the unitarily implemented channels
$$
\Phi_m\left(  B\right)  =U^{\dagger}_mBU_m,\;\Psi\left(  B\right)  =V^{\dagger
}_mBV_m
$$
with the unitaries $U_m,V_m$ chosen in such a way that $U_m\neq V_m$
but
$$
\lim_{m \rightarrow \infty}
  \frac{1}{m}\Tr(U^{\dagger}_m V_m)=1.
$$
Thus, the channel fidelity
  (\ref{eq:chfid}), apart from being applicable only in
finite-dimensional settings, has the distinct disadvantage of not
  being equivalent to the cb-norm distance, in contrast to the case of
  the Uhlmann fidelity (\ref{eq:statefid}) and the trace-norm distance
  (\ref{eq:tnd}) on the state space of a quantum-mechanical system.

The goal of this paper is to define and study a new fidelity criterion
for
quantum channels, which we term the {\em minimax fidelity} and which is a
noncommutative generalization of maximal Hellinger distance between
two positive kernels in classical probability theory. Unlike the
channel fidelity (\ref{eq:chfid}) of Ref.~\cite{Rag01}, the minimax
fidelity is not only well-defined for channels between
finite-dimensional algebras, but also applies to a certain class of
channels between infinite-dimensional algebras (explicitly, those
channels that possess an operator-valued Radon--Nikodym density with
respect to the trace in the sense of
Belavkin--Staszewski \cite{BeSta86}) and is equivalent to the CB-norm
distance, echoing the way the Uhlmann fidelity
(\ref{eq:statefid}) for density operators is equivalent to the
trace-norm distance (\ref{eq:tnd}).

Apart from these technical features, the minimax fidelity
$f(\Phi,\Psi)$ between two quantum channels $\Phi,\Psi$ has a direct {\em operational} meaning: intuitively, it is
defined as the minimum overlap of output states (density operators) of
the predual channels $\Phi_\intercal,\Psi_\intercal$, when the operator-sum decompositions \cite{NC00} of
the latter are chosen to be maximally overlapping; this is spelled out
in precise terms in Section~\ref{ssec:qofid_kraus}. Our
central result (Theorem~\ref{thm:mmfid}) demonstrates that the minimax
fidelity is independent of the order of these two
optimizations. Furthermore, the equivalence of our minimax fidelity to
the CB-norm distance, which is stated precisely in
Section~\ref{sec:properties} in terms of {\em dimension-free} bounds,
is a promising avenue for the study and characterization of
dimension-free bounds (whenever they exist) on other operationally meaningful distance
measures for quantum operations \cite{GLN04} in terms of the CB-norm
distance. As pointed out in Ref.~\cite{KreWer04}, such bounds are
crucial for a successful generalization of the usual quantum capacity of a
channel \cite{Key02,NC00} (i.e., with respect to the identity channel)
to the case of comparing quantum channels to an arbitrary reference
channel. We plan to pursue these matters further in a future publication.

The paper is organized as follows. In Section~\ref{sec:prelims} we fix
the definitions and notation used throughout the
paper. The minimax fidelity is then introduced in
Section~\ref{sec:opdist}. Section~\ref{sec:evaluate} is devoted to the
evaluation of the minimax fidelities in the various mathematical settings that arise
in quantum information theory. Next, in Section~\ref{sec:properties},
we list key properties of the minimax
fidelity. Finally, in Section~\ref{sec:ex} we sketch some example
applications of the minimax fidelity to several problems of quantum information
theory.

\section{Preliminaries, definitions, notation}
\label{sec:prelims}

\subsection{Pairings, states, operations}
\label{ssec:pairing}

Let $\fh$ be a complex separable Hilbert space; let $\cB$ denote
the Banach algebra $\cB(\fh)$ of all bounded linear operators on $\fh$
with the usual operator norm $\norm{\bullet}$; and let $\cB_\intercal$ denote the Banach space
$\cB_\intercal(\fh)$ of trace-class operators on $\fh$ with the trace norm $\trnorm{\bullet}$. The set of normal states on $\cB$, i.e., ultraweakly continuous positive unital
linear functionals on $\cB$, will be denoted by $\cS(\cB)$ or,
whenever we need to exhibit the underlying Hilbert space explicitly,
by $\cS(\fh)$. Generic elements of $\cS(\cB)$ will be denoted by the
stylized Greek letters $\varpi,\varrho,\varsigma$. Note that the
operator norm on $\cB$ can be written as $\norm{B} = \sup\big\{\varrho
(\abs{B}) : \varrho \in \cS(\cB)\big\}$.

We equip $\fh$ (and shall equip all Hilbert spaces introduced in the sequel)
with an isometric involution $J = J^\dagger$, $J^2 = \idty_\fh$,
having the properties of complex conjugation,
\[
J \sum_j \lambda_j \eta_j = \sum_j \overline{\lambda_j}J\eta_j, \qquad
\forall \lambda_j \in \C,\,\eta_j \in \fh.
\]
We can thus define the {\em transpose} of any $B \in \cB$ as $\tran{B}
:= J B^\dagger J$, as well as introduce the trace pairing
\cite{BeOhy01}
\begin{equation}
(B,\rho) := \Tr(B\tran{\rho}) = \Tr(\tran{B}\rho), \qquad \forall B \in \cB,\,\rho
  \in \cB_\intercal
\label{qpairing}
\end{equation}
of $\cB$ and $\cB_\intercal$. Under this pairing, which differs from
the usual one in that $B \in \cB$ is paired with the
transpose of $\rho \in
\cB_\intercal$ rather than directly with $\rho$, normal linear
functionals on $\cB$ are in a one-to-one correspondence with the
elements of $\cB_\intercal$. Thus to each normal state
$\varrho$ we associate a unique positive trace-class operator with unit trace, denoted by the standard Greek
letter $\rho$ and referred to as the {\em density operator}
corresponding to $\varrho$, via $\varrho(B) = (B,\rho)$ for all $B \in
\cB$. Similarly, density operators corresponding to states denoted by $\varpi$
and $\varsigma$ will be denoted by $\pi$ and $\sigma$
respectively.

Apart from natural arguments from standard representation theory of
operator algebras, one reason why we chose to pair $B$ with
the transposed operator $\tran{\rho} = J\rho^\dag J$, rather than with
$\rho$, is to be able to keep all notations conveniently parallel to the classical
(commutative) case, as will be amply demonstrated throughout the
paper. Note also that we can fix a complete orthonormal
basis $\{ \ket{j} \}$ of $\fh$ and express the pairing
(\ref{qpairing}) in terms of the matrix elements of $B$ and $\rho$ as
\[
(B,\rho) = \sum_{j,k} \bra{j}B\ket{k} \cdot \bra{j}\rho\ket{k} \equiv
\sum_{j,k} B_{jk}\rho^{jk},
\]
where we have used the covariant indices for the matrix elements of
bounded operators in $\cB$ and the contravariant indices for the matrix
elements of trace-class operators in $\cB_\intercal$, when the latter
are identified via the pairing (\ref{qpairing}) with
normal linear functionals on $\cB$. Yet another reason to opt for the
pairing of $B$ with the transposed operator $\tran{\rho}$, further elaborated upon in Section~\ref{ssec:opdens}, is that then
the density operator $\rho$ of a normal state $\varrho$ will coincide
with the operational density of $\varrho$, understood as a quantum
operation from $\cB$ into the Abelian algebra $\C$.

Introducing another Hilbert space $\fg$, the algebra $\cA := \cB(\fg)$
and the trace class $\cA_\intercal := \cB_\intercal(\fg)$, let us
consider {\em quantum operations}, i.e., the completely positive
normal linear mappings $\map{\Phi}{\cB}{\cA}$ such that
$\Phi(\idty_\fh) \le \idty_\fg$; if $\Phi(\idty_\fh) = \idty_\fg$,
then $\Phi$ is referred to as a {\em quantum channel}. Any quantum
operation $\Phi$ possesses a unique {\em
  predual} $\map{\Phi_\intercal}{\cA_\intercal}{\cB_\intercal}$,
defined as the transpose of $\Phi$ with respect to the trace pairing
(\ref{qpairing}), i.e.,
\begin{equation}
\big(\Phi(B),\rho\big) = \big(B,\Phi_\intercal(\rho)\big), \qquad
\forall B \in \cB,\, \rho \in \cA_\intercal.
\label{eq:predual}
\end{equation}
Conversely, given a normal completely positive linear map
$\map{\Phi}{\cA_\intercal}{\cB_\intercal}$ such that $\Tr_\fh
\Phi(\rho) \le \Tr_\fg \rho$ for all $\rho \in \cA_\intercal$, we define its {\em dual}
with respect to the trace pairing (\ref{qpairing}) as the unique
mapping $\map{\Phi^\intercal}{\cB}{\cA}$ for which
\begin{equation}
\big(B,\Phi(\rho)\big) = \big(\Phi^\intercal(B),\rho\big), \qquad
\forall B \in \cB,\, \rho \in \cA_\intercal.
\label{eq:dual}
\end{equation}
Using these definitions, one readily obtains that
$\Phi^\intercal_\intercal = \Phi$ for any normal completely positive
map $\map{\Phi}{\cB}{\cA}$. Alternatively, one may define the predual
of a normal completely positive map $\map{\Phi}{\cB}{\cA}$ as the
unique normal completely positive map
$\map{\Phi_\intercal}{\cA_\intercal}{\cB_\intercal}$ such that
$\Phi^\intercal_\intercal = \Phi$.

If $\Phi$ is given in the Kraus form \cite{Kra83} $\Phi\left(  B\right)  =\sum
F_{j}^{\dagger}BF_{j}$, or more generally as an integral
\begin{equation}
\Phi\left(  B\right)  =\int_{Z}F\left(  z\right)  ^{\dagger}BF\left(
z\right)  \mathrm{d}\mu\left(  z\right),  \label{eq:genkraus}
\end{equation}
with respect to a positive measure $\mu$ on a measurable space $(Z,\cB_Z)$,
where the integration is understood in the sense of Bochner \cite{Yos65}, then the
predual map $\Phi_\intercal$ has the transposed integral form
\[
\Phi_{\intercal}\left(  \rho\right)  =\int_{Z}F_{\intercal}\left(  z\right)
^{\dagger}\rho F_{\intercal}\left(  z\right)  \mathrm{d}\mu\left(  z\right)  ,
\]
where $\fg\ni\xi\longmapsto\bra{\xi} F_{\intercal}\left(  z\right)  $ are
Hilbert-transposed to the operators $\fh\ni\eta\longmapsto\bra{\eta} F\left(
z\right)  $, that is $F_{\intercal}\left(  z\right)  =\tran{F\left(
  z\right)}$ for all $z \in Z$.

Any normal state $\varrho \in \cS(\cB)$ is automatically a quantum
channel from $\cB$ into the Abelian algebra $\C$, and it is readily
seen that the density operator $\rho$ of $\varrho$, understood as acting on
$\lambda \in \C$ on the right, $\C \ni \lambda \longmapsto \lambda \rho$, is precisely the predual
$\map{\varrho_\intercal}{\C}{\cB_\intercal}$. Indeed, given $B \in \cB$ and $\lambda \in \C$,
we have
\[
\big(\varrho(B),\lambda\big) = (B,\lambda \rho) = \big(B,\varrho_\intercal(\lambda)\big),
\]
which proves our claim that $\rho = \varrho_\intercal$. Thus we also
have that $\varrho = \varrho^\intercal_\intercal = \rho^\intercal$.

\subsection{Operational densities}
\label{ssec:opdens}

In order to avoid technicalities involving unbounded
operators, we shall henceforth assume that all quantum operations we deal
with are \emph{completely majorized} by the trace, considered as the
map $\tau\left(  \sigma\right) =\openone_{\fg}\Tr \sigma$ of $\cB_{\intercal
}$ into $\cA=\cB\left(  \fg\right)  $, in the sense
\cite{BeSta86} that there exists a constant $\lambda>0$ such that the difference
$\lambda\tau-\Phi$ is a completely positive
map $\cB_{\intercal}\longrightarrow \cA$. For example, this condition
is satisfied by all quantum operations between finite-dimensional
algebras \cite{Rag03}. As was proven in
\cite{BeSta86}, in this case there exists a unique positive operator
$\Phi_\tau$ on the Hilbert
space $\mathcal{H}:=\fg\otimes\fh$, called the {\em density} of $\Phi$
with respect to the trace $\tau$, such that
\begin{equation}
\Phi(B) = \Tr_\fh \big[  ( \idty_\fg \otimes \tran{B})  \Phi_\tau\big],
\label{eq:cptran}
\end{equation}
where $\Tr_\fh Y$, $Y \in \cB(\cH)$, denotes the partial trace of $Y$
with respect to $\fh$,
$$
(\Tr_\fh Y, \rho) = \big(Y, \rho \tp \idty_\fh  \big), \qquad \forall
\rho \in \cB_\intercal(\fg).
$$
Moreover, $\Phi_\tau$ as a linear
operator on $\cH$ is bounded and majorized by $\lambda$:
$0\leq\Phi_\tau\le \lambda \idty_\cH$, and the operation
is unital, $\Phi(\idty_\fh)=\idty_\fg$ [contractive, $\Phi( \idty_\fh
  ) \le \idty_\fg$] if and only if $\Tr_\fh\Phi_\tau=\idty_\fg$
($\Tr_\fh \Phi_\tau \le \idty_\fg$). This is equivalent to saying
that the predual map
$\map{\Phi_\intercal}{\cA_\intercal}{\cB_\intercal}$, which, using
Eqs.~(\ref{eq:predual}) and (\ref{eq:cptran}), can be written as
\begin{equation}
\Phi_{\intercal}\left(  \rho\right)  =\Tr_\fg\big[
\Phi_\tau (  \tran{\rho} \otimes \idty_\fh) \big],
\label{eq:cptranpredual}
\end{equation}
is trace-preserving (trace-decreasing).

As an example, consider a normal state $\varrho$ on $\cB$, which,
being a quantum channel
into $\C$, satisfies the complete majorization condition with $\lambda
= \| \rho \|$, where $\rho$ is the density operator of
$\varrho$. Furthermore, it is easy to see that $\varrho_\tau =
\rho$. Indeed, we can write
$$
\varrho(B) = (B,\rho) = \Tr (B\tran{\rho}) = \Tr (\tran{B}\rho) = \Tr_\fh \big[(\idty_\C \tp \tran{B})\rho],
$$
and the desired result follows upon comparing this with Eq.~(\ref{eq:cptran}). This provides additional justification for our definition of
the trace pairing in Eq.~(\ref{qpairing}), since we then have that
$\varrho_\intercal = \rho = \varrho_\tau$ for any normal state $\varrho$.

If the operation $\map{\Phi}{\cB}{\cA}$ is given in the generalized
Kraus form (\ref{eq:genkraus}), we can write down its operational
density $\Phi_\tau$ explicitly. To this end, suppose that all operators $F\left(  z\right)
$ are determined by generalized bra-vectors $\Gamma\left(  z\right)
=(F\left(  z\right)  |$, densely defined as the linear functionals
\[
\Gamma(  z)  \ket{\xi\otimes\eta} = \bra{\xi} F(z)
\ket{\eta} \equiv \big(F(z)  \big|\ket{\xi\otimes\eta}\big)
\]
on the linear span of the ket-vectors
$\ket{\xi\otimes\eta} \equiv \tran{\xi\otimes\eta}$ in
$\mathcal{H}=\fg\otimes\fh$, where $\xi\in\fg$ is also treated as a
bra-vector such that $J\xi = \bra{\xi}$ and $\ket{\xi} = \tran{\xi}$. Then the operational density
$\Phi_\tau$ of $\Phi$ is given by the corresponding decomposition
\begin{equation}
\Phi_\tau=\int\Gamma\left(  z\right)  ^{\dagger}\Gamma\left(
z\right)  \mathrm{d}\mu\left(  z\right)  \equiv\Gamma^{\dagger}
\Gamma,\label{qdecomp}
\end{equation}
where the integral is, again, understood in the sense of Bochner.

\subsection{Completely bounded maps}
\label{ssec:cbmaps}

Completely positive linear maps between operator algebras are a
special case of {\em completely bounded} maps \cite{Pau03}. Consider, as before,
the algebras $\cB = \cB(\fh)$ and
$\cA = \cB(\fg)$. For each $n \in \N$ define
the $n$th {\em matrix level} $\cM_n(\cB)\simeq
\cB \otimes \cM_n$, where $\cM_n$ denotes the
algebra of $n\times n$ matrices with complex entries. That is,
$\cM_n(\cB)$ is the space of $n\times n$ matrices with
$\cB$-valued entries,
$$
\cM_n(\cB) := \left\{
[B_{ij}] : B_{ij} \in \cB, 1\le i,j \le n\right\}.
$$
Analogous
construction can also be applied to $\cA$ to yield the matrix
levels $\cM_n(\cA)$. Each matrix level $\cM_n(\cB)$
inherits a $*$-algebra structure from $\cB$ through
$$
[B_{ij}] [C_{ij}] := \left[ \sum^n_{k=1} B_{ik} C_{kj}\right],\;
[B_{ij}]^\dag := [B^\dag_{ij}].
$$
In fact, by identifying $\cM_n(\cB)$ via a natural $*$-isomorphism with the algebra
$\cB(\fh^{(n)})$ of bounded linear operators on
$\fh^{(n)}$, the direct sum of $n$ copies of $\fh$,
one can make $\cM_n(\cB)$ into a C*-algebra. Thus,
each matrix level of $\cB$ possesses a unique C*-norm.

Now, for any $n \in \N$ a linear
map $\Lambda : \cB \longrightarrow \cA$ induces the map
$\Lambda^{(n)} := \Lambda \otimes \mathrm{id}_n$ from
$\cM_n(\cB)$ into $\cM_n(\cA)$,
defined by $\Lambda^{(n)} : [B_{ij}] \longmapsto [\Lambda(B_{ij})]$. Let
us define the {\em norm of complete boundedness} (or CB-norm) by $\cbnorm{\Lambda} := \sup \left\{ \norm{\Lambda^{(n)}} : n \in
    \N \right\}$, where
$$
\norm{\Lambda^{(n)}} := \sup_{B \in \cM_n(\cB),
    \norm{B} \le 1} 
    \norm{\Lambda^{(n)}(B)}
$$
is the usual operator norm of $\Lambda^{(n)}$. A linear map $\Lambda :
\cB \longrightarrow \cA$ is called {\em completely bounded} if
$\cbnorm{\Lambda} < \infty$. Every completely positive map $\Phi :
\cB \longrightarrow \cA$ is automatically completely bounded, with
$\cbnorm{\Phi} = \norm{\Phi(\idty_\fh)}$. For a general
completely bounded map $\Lambda$, one has, by definition, $\|
\Lambda(\idty_\fh)\| \le \| \Lambda \| \le \|
\Lambda \|_\mathrm{cb}$.

Passing to the predual map $\Lambda_\intercal  :
\cA_\intercal \longrightarrow \cB_\intercal$, we can similarly
define induced maps $\Lambda^{(n)}_\intercal :
\cM_n(\cA_\intercal) \longrightarrow \cM_n(\cB_\intercal)$, $n \in \mathbb{N}$, and the predual CB-norm
$$
\| \Lambda \|^\intercal_{\rm cb} := \sup_{n \in \mathbb{N}}
\| \Lambda^{(n)} \|_\intercal,
$$
where
$$
\| \Lambda^{(n)} \|_\intercal := \sup_{\rho \in
  \cM_n(\cA_\intercal) : \| \rho \|_\intercal
  \le 1} \| \Lambda^{(n)}(\rho) \|_\intercal.
$$
It is easy to see that $\| \Lambda^{(n)} \| = \|
\Lambda^{(n)}_\intercal \|_\intercal$ for all $n \in \mathbb{N}$,
so that $\| \Lambda \|_{\rm cb} = \| \Lambda_\intercal
\|^\intercal_{\rm cb}$. It is also straightforward to see that the
``unstabilized'' norms $\| \bullet \|$ and $\| \bullet
\|_\intercal$ are tensor-supermultiplicative (i.e., $\| \Lambda_1
\otimes \Lambda_2 \| \ge \| \Lambda_1 \| \| \Lambda_2
\|$), whereas the corresponding CB-norms are tensor-multiplicative
(i.e., $\| \Lambda_1 \otimes \Lambda_2 \|_{\rm cb} = \|
\Lambda_1 \|_{\rm cb} \| \Lambda_2 \|_{\rm cb}$).

There is also a useful non-variational formula for the CB-norm of a
map $\map{\Lambda}{\cB}{\cA}$. Namely, let $\ell^2$ denote the Hilbert
space of square-summable infinite sequences of complex numbers, and
let $\cK(\ell^2)$ denote the space of compact operators on
$\ell^2$. Then $\cbnorm{\Lambda} = \norm{\Lambda \tp
  \id_{\cK(\ell^2)}}$. Since we have assumed that $\cB = \cB(\fh)$
with $\fh$ a complex separable Hilbert space, and since all complex
separable Hilbert spaces are canonically isomorphic to $\ell^2$, we may also write
$\cbnorm{\Lambda} = \norm{\Lambda \tp \id_{\cK(\fh)}}$.

\subsection{Miscellany}
\label{ssec:misc}

Any positive operator $B \in \cB(\fh)$ has a unique positive square
root, denoted by $B^{1/2}$ and defined as the positive operator $X \in \cB(\fh)$ such that
$B=X^2$. This definition can be extended to any operator $A$ that is
similar to a positive operator $\Delta \in \cB(\fh)$, in the
sense that there exists an operator $S \in \cB(\fh)$ such that $A =
S\Delta S^+$, where $S^+$ is the pseudoinverse of $S$, equal to
$S^{-1}$ on $\ran S$ and
to $0$ on $\ker S$. In that case, we may {\em define} $\sqrt{A} :=
S\Delta^{1/2}S^+$. From now on, in order to distinguish this extended
definition of the square root from the usual one, we shall always use
the square root symbol $\sqrt{\bullet}$ for this extended definition,
and reserve the exponent notation $\bullet^{1/2}$ for the usual definition.

Consider now two positive operators $A,B \in \cB(\fh)$. It is easy to see
 that their product $AB$ is similar to $A^{1/2}B A^{1/2}$ with $S =
 A^{1/2}$. Note that the operator $AB$ is positive when restricted to
the closure of $\ran A$, when the latter is equipped with the weighted inner
product $\braket{\upsilon}{\chi}_A :=
\braket{A^{-1/2}\upsilon}{A^{-1/2}\chi}$:
$$
\braket{\upsilon}{AB\upsilon}_A =
\braket{A^{-1/2}\upsilon}{A^{-1/2}AB\upsilon} =
  \braket{\upsilon}{B\upsilon} \ge 0, \qquad \forall \upsilon \in
  \overline{\ran A}.
$$
Thus we may define $\sqrt{AB} := S (A^{1/2}B A^{1/2})S^+$ with $S = A^{1/2}$.

This notation, again, allows for a convenient parallelism between the classical (commutative) formalism
and the quantum (noncommutative) one. Indeed, consider two mutually
commuting positive trace-class operators $\rho,\sigma$, let $\{
\ket{x} \}$ denote the set of their common eigenvectors, and let
$\rho_x \equiv \bra{x}\rho\ket{x},\sigma_x \equiv \bra{x}\sigma\ket{x}$ denote the corresponding eigenvalues. Then
$\sqrt{\rho \sigma}$ is also trace-class, and
$$
\Tr \sqrt{\rho \sigma} = \sum_x \sqrt{\rho_x\sigma_x}.
$$
If $\Tr \rho = 1 = \Tr \sigma$, then $P:=\{\rho_x\}$ and $Q:=\{\sigma_x\}$
are probability distributions, and $\Tr \sqrt{\rho \sigma}$ then gives
the classical fidelity (also known as the {\em Bhattacharyya
  coefficient}) \cite{FG99} $F(P,Q)$ between $P$ and $Q$.

Our main technical tool in this paper is given by the following:

\begin{lemma}\label{lm:lagrange}
Let $\cH$ be a complex separable Hilbert space, and let $R,S \in
\cB(\cH)$ be positive operators such that $R^{1/2}SR^{1/2}$ is
trace-class. Then the supremum
\begin{equation}
\sup_{X,Y\in\cB(\cH)}\left\{  \Tr (  X^\dag Y+ Y^\dag X)
:X^\dag X = R,Y^\dag Y = S\right\} = 2\Tr\sqrt{RS}
\label{lemsup}
\end{equation}
is achieved on any $X\in\cB(\cH)$ satisfying
the condition $X^{\dagger}X=R$, say $X=R^{1/2}$, and $Y=Y_\opt$
satisfying the equation
\begin{equation}
Y_\opt X^\dag = (XSX^\dag)^{1/2} = XY_\opt ^{\dagger}.
\label{lagequ}
\end{equation}
\end{lemma}

\begin{proof}
To prove the lemma one can use either the polar decomposition or the method of
Lagrange multipliers. We shall use the latter. Fixing an $X$ satisfying
$X^{\dagger}X=R$, we can write the Lagrange function as
\[
{\sf L}=\Tr ( X^\dag Y + Y^\dag X- Y^\dag YL),
\]
where $L=L^\dag \in \cB(\cH)$ is the operator-valued Lagrange
multiplier corresponding to the hermiticity condition $S=Y^\dag Y =
S^\dag$. At the stationary point 
\[
\delta{\sf L}=\Tr( X^\dag - LY^\dag) \delta Y + (X-YL)\delta Y^\dag = 0,
\]
so $Y=Y_\opt$ must satisfy the equation $YL=X$ (the other equation,
$LY^\dag = X^\dag$, corresponding to $Y^\dag = Y^\dag_\opt$, is obtained by taking the Hermitian adjoint). Thus $Y_\opt =XL^{-1}$, where $L^{-1}$
should be determined from $L^{-1}X^\dag XL^{-1} = S$. Multiplying this on
the left by $X$ and on the right by $X^{\dagger}$ yields
$(XL^{-1}X^\dag)^2 = XSX^\dag$, or $XL^{-1}X^\dag =
(XSX^\dag)^{1/2}$. Thus, we indeed have that $Y_\opt X^\dag =
(XSX^\dag)^{1/2} = XY^\dag_\opt$, and therefore that
\begin{equation}
\Tr ( Y_\opt X^\dag + XY_\opt ^\dag) = 2\Tr\big((XSX^\dag)^{1/2}\big).
\label{lemproofsup}
\end{equation}
This extremal value is precisely the  maximal value due to convexity
of the function being maximized in Eq.~(\ref{lemsup}). Note that,
since $(U^\dag X S X^\dag U)^{1/2} = U^\dag (XSX^\dag)^{1/2}U$ for any unitary $U$, the
value of the supremum in Eq.~(\ref{lemsup}), which coincides with Eq.~(\ref{lemproofsup}), does not depend
on the choice of $X$ satisfying $X^\dag X = R$. Indeed, by virtue of the polar
decomposition $X=UR^{1/2}$,
$$
2\Tr(XSX^\dag)^{1/2} = 2 \Tr\big( U^\dag (XSX^\dag)^{1/2}U \big)  = 2
\Tr \big(
(R^{1/2}SR^{1/2})^{1/2}\big).
$$
Rewriting this trace in the equivalent form $2\Tr (X^\dag Y_\opt)$ with
$$
X^\dag Y_\opt = R^{1/2}(R^{1/2}SR^{1/2})^{1/2}R^{-1/2} \equiv
\sqrt{RS}
$$
corresponding to $X=R^{1/2}$, we obtain the extremal value in
Eq.~(\ref{lemsup}).   
\end{proof}

We shall also need the following simple, but useful, result:

\begin{lemma}\label{lm:homf} Let $S$ be a compact subset of a complex Banach space $V$, such that
  $x \in S$ implies $\lambda x \in S$ for all $\lambda \in \C$ with
  $\abs{\lambda}=1$. Let $\map{f}{V}{\C}$ be a continuous function which is
  homogeneous of order $1$, i.e., $f(\lambda x) = \lambda f(x)$ for all
  $\lambda \in \C$ and all $x \in V$. Then
\begin{equation}
\sup_{x \in S}\abs{f(x)} = \sup_{x \in S}\Re f(x).
\label{eq:tropt}
\end{equation}
\end{lemma}

\begin{proof} Let $x^* \in S$ be such that $\abs{f(x^*)} = \sup_{x
    \in S} \abs{f(x)}$, with $f(x^*) =
    \abs{f(x^*)}\mathrm{e}^{\mathrm{i}\arg f(x^*)}$. Let $x^{**} :=
    \mathrm{e}^{-\mathrm{i}\arg{f(x^*)}}x^*$. By the homogeneity of $f$,
$$
f(x^{**}) = \mathrm{e}^{-\mathrm{i}\arg{f(x^*)}}f(x^*) = \abs{f(x^*)}.
$$
But then $\abs{f(x^{**})} = f(x^{**}) = \Re 
f(x^{**})$. Since $\Re \lambda \le \abs{\lambda}$ for all
$\lambda \in \C$, the lemma is proved.
\end{proof}

\section{Operational fidelities and distances}
\label{sec:opdist}

\subsection{Classical kernel fidelity}
\label{ssec:kernels}

The fidelity distinguishing different quantum operations without the
restriction on the Hilbert space dimensionality was suggested by Belavkin
in Ref.~\cite{Be02} on the basis of a noncommutative generalization of
the maximal Hellinger distance between two positive kernels. Namely, given
a locally compact space $X$ and a measure space $(Y,\cB_Y,\mu)$, where $\mu$ is a
positive measure, let us denote by $\cA$ the
algebra $\mathcal{C}(X)$ of bounded continuous functions on $X$, and
by $\cB_\intercal \equiv \mathcal{C}_\intercal(Y)$ the space of absolutely $\mu$-integrable complex
functions on $Y$. A {\em positive kernel} $\mathbf{P}$ is then given in
terms of a function $\map{p(\bullet|\bullet)}{Y \times X}{\R^+}$, such
that $P_x := p(\bullet|x) \in \cB_\intercal$ for all $x \in X$,
while $P := \int_Y p(y|\bullet)\mathrm{d}\mu(y) \in \cA$. Given
two positive kernels $\mathbf{P}$ and $\mathbf{Q}$, the squared
pointwise Hellinger
distance
\begin{eqnarray}
d^2_H(P_x,Q_x) &:=& \frac{1}{2}\int\left(\sqrt{p(y|x)} -
\sqrt{q(y|x)}\right)^2\mathrm{d}\mu(y)\nonumber\\
&=& \int\left[\frac{1}{2}\big(p(y|x) + q(y|x)\big) - \sqrt{p(y|x)q(y|x)}\right]\mathrm{d}\mu(y)
\label{eq:ptwhellinger}
\end{eqnarray}
is well-defined and finite for each $x \in X$, so that we can define
\begin{equation}
d^2_H(\boldsymbol{P},\boldsymbol{Q}) := \frac{1}{2}\sup_{x \in X} \int\left(
\sqrt{p(y|x)}-\sqrt{q(y|x)}\right)^2 \mathrm{d}\mu(y)  
\equiv \|d^2_H(P_x, Q_x)\|,
\label{ccfdist}
\end{equation}
the last expression indicating the fact that
$d^2_H(\boldsymbol{P},\boldsymbol{Q})$ is given by the supremum of the squared pointwise Hellinger distance
(\ref{eq:ptwhellinger}) over all $x \in X$. Note that the squared Hellinger
distance $d^2_H(P,Q)$ between two positive distributions
$P=p(\bullet)$ and $Q=q(\bullet)$ is the minimal mean
quadratic distance
\begin{align}
d^2_H(P,Q) & = \frac{1}{2}\inf_{\chi,\psi \in \cC(Y)} \left\{
\int\abs{\chi(y) - \psi(y)}^2\mathrm{d}\mu(y) : \abs{\chi(\bullet)}^2 = p(\bullet),\abs{\psi(\bullet)}^2 =
q(\bullet)\right\}\nonumber \\
& =\left( 1,\frac{1}{2}(P+Q) \right) - \sup_{\psi:\abs{\psi(\bullet)}^2
= q(\bullet)} \int\sqrt{p(y)}\Re \psi(y) \mathrm{d}\mu(y), \label{cfdist}
\end{align}
where $( f,P ) =\int f(y) p(y) \mathrm{d}\mu(y)$ denotes the integral pairing of
$f \in \mathcal{C}(Y)$ with $P \in \mathcal{C}_\intercal(Y)$. The {\em
  relative fidelity}
\begin{eqnarray}
f(P,Q) &=& \frac{1}{\sqrt{(1,P)(1,Q)}}
\sup_{\psi:\abs{\psi(\bullet)}^2 = q(\bullet)}
\int \sqrt{p(y)}\Re \psi(y) \mathrm{d}\mu(y) \nonumber \\
&=&
\frac{(1,\sqrt{PQ})}{\sqrt{(1,P)(1,Q) }},
\label{cfid}
\end{eqnarray}
of the distributions $P$ and $Q$ is
obviously related to the distance (\ref{cfdist}) by
\begin{equation}
d^2_H(P,Q) + \sqrt{(1,P)(1,Q) } f(P,Q) =
\left(1,\frac{1}{2}(P+Q)\right).
\label{cdisfid}
\end{equation}
If $P_x := p(\bullet|x)$ and $Q_x := q(\bullet|x)$ are conditional
distributions with constant integrals $( 1,P_x)$ and $( 1,Q_x ) $,
e.g., normalized to unity, this relation also remains valid for the minimal
fidelity
$$
f(\boldsymbol{P},\boldsymbol{Q}) = \inf_{x \in X}f(P_x, Q_x),
$$
which can alternatively be defined by the minimax formula
\begin{equation}
f(\boldsymbol{P},\boldsymbol{Q}) = \inf_{x \in X}\sup_{\psi:
\abs{\psi(\bullet|x)}^2 = Q_x(\bullet)}
\frac{(1,\sqrt{P_x }\Re \psi(\bullet|x))}{\sqrt{(1,P_x)(1,Q_x)}},
\label{ccofid}
\end{equation}
where the supremum is achieved on $\psi(\bullet|x) \equiv
\psi_\opt(\bullet|x)$ satisfying $\psi (y|x) = \sqrt{q(y|x)}$. In
particular, if $\boldsymbol{P}$ and $\boldsymbol{Q}$ are probability
kernels, $(1,P_x) = 1 = (1,Q_x)$ for all $x \in X$, then
$$
d^2_H(\boldsymbol{P},\boldsymbol{Q}) = 1 - \inf_{x \in X}
\int\sqrt{p(y|x)q(y|x)} \mathrm{d}\mu(y)
\equiv 1 - f(\boldsymbol{P},\boldsymbol{Q}),
$$
where
\begin{equation}
f(\boldsymbol{P},\boldsymbol{Q}) = \inf_{x \in X} \int\sqrt{p(y|x)
q(y|x)} \mathrm{d}\mu(y)
\equiv \inf_{x \in X}(1,\sqrt{P_x Q_x})
\label{cccfid}
\end{equation}
is the minimax fidelity of the classical channels described by these kernels.

\subsection{Quantum operational fidelity}
\label{ssec:opfid}

Generalizing Eq.~(\ref{ccfdist}), one can define the squared Hellinger distance between quantum operations $\Phi$ and
$\Psi$ with the respective operational densities $\Phi_\tau,\Psi_\tau \in
\cB(\cH)$, $\cH = \fg \tp \fh$, as
\begin{equation}
d^2_H (\Phi,\Psi) = \frac{1}{2} \inf_{\Gamma,\Upsilon \in \cB(\cH)} 
\left\{ \Norm{ \Tr_\fh (\Gamma-\Upsilon)^\dag
(\Gamma-\Upsilon) } : \Gamma^\dag \Gamma = \Phi_\tau,
\Upsilon^\dag \Upsilon = \Psi_\tau \right\}.
\label{qqfdist}
\end{equation}
The operators $\Gamma,\Upsilon \in
\cB(\cH)$, such that $\Gamma^\dag \Gamma = \Phi_\tau$ and
$\Upsilon^\dag \Upsilon = \Psi_\tau$, are naturally thought of as the {\em
purifications} of $\Phi_\tau$ and $\Psi_\tau$, respectively. This
means that we can fix an orthonormal basis $\{ \ket{j} \}$ of $\cH$,
say the product basis
$\ket{j} = \ket{i} \otimes \ket{k} \equiv \ket{i,k}$, where
$\{\ket{i}\}$ and $\{\ket{k}\}$ are some fixed orthonormal bases of $\fg$ and
$\fh$ respectively, and represent any such $\Gamma$ and $\Upsilon$ as
strongly convergent sums
\begin{equation}
\Gamma = \sum_j \ketbra{j}{j}\Gamma \equiv \sum_j \ket{j} (F_j|, \;
\Upsilon = \sum_j \ketbra{j}{j}\Upsilon \equiv \sum_j \ket{j}(V_{j}|,
\label{eq:kraus}
\end{equation}
where the generalized bra-vectors $(F_j|$ define the bounded
operators $\map{F_j, V_j}{\fg}{\fh}$ through
$$
\bra{k} F_j \ket{i} = (F_j|(\ket{i} \otimes \ket{k})
 = \bra{j} \Gamma \ket{i,k} ,\;
\bra{k} V_j \ket{i} = (V_j|(\ket{i} \otimes \ket{k}) = \bra{j}
 \Upsilon \ket{i,k}.
$$
As seen directly from this definition, the mapping $(F| \longmapsto F$
is linear: $(a F + b G| \longmapsto aF + bG$. Using Eq.~(\ref{eq:kraus}), we may write
\begin{equation}
\Phi_\tau=\sum_j |F_{j})(F_{j}| \equiv \Gamma^ \dag\Gamma, \;
\Psi_\tau=\sum_j |V_{j})(V_{j}| \equiv \Upsilon^\dag \Upsilon,
\label{qqpurif}
\end{equation}
where the sums converge in the strong operator topology. This
determines the Kraus decompositions \cite{Kra83} $\Phi (B) = \sum_j
F_j^\dag B F_j$, $\Psi (B) = \sum_j V_j^\dag B V_j$ of the maps
$\map{\Phi,\Psi}{\cB}{\cA}$. Analogously, upon defining the mappings $\map{F,V}{\fg}{\fh \otimes \cH}$ by
$$
F \upsilon := \sum_j F_j \upsilon \otimes \ket{j}, \;
V \upsilon := \sum_j V_j \upsilon \otimes \ket{j},
$$
we can write the maps $\Phi,\Psi$ in the Stinespring form \cite{Sti55} as $\Phi (B) =
F^\dag (B \otimes \openone_\mathcal{H}) F$ and $\Psi (B) = V^\dag (B \otimes \openone_\mathcal{H})V$.

Taking into account the fact that $\| A^{\dagger}A\| =\sup_{\varrho
\in\mathcal{S}\left(  \fg\right)  }\varrho(A^\dagger A)$ and defining
the positive function
\begin{eqnarray*}
&& \map{c(\bullet;\bullet)}{\cB(\cH)\times\cB_\intercal(\fg)}{\R}\\
&& c( A;\rho ) := \frac{1}{2}\Tr \left (A(\rho \otimes \idty_\fh)
A^\dag \right),
\end{eqnarray*}
we can rewrite the fidelity distance (\ref{qqfdist}) in the following minimax form:
\begin{equation}
d^2_H(\Phi,\Psi) = \inf_{\Gamma,\Upsilon \in \cB(\cH)} \left\{
\sup_{\varrho \in \cS(\fg)} c(\Gamma-\Upsilon;\rho) : \Gamma^\dag \Gamma = \Phi_\tau, \Upsilon^\dag \Upsilon =
\Psi_\tau \right\}.
\label{qqmmdist}
\end{equation}
On the other hand, generalizing Eq.~(\ref{eq:ptwhellinger}) to quantum operations, we can define the squared pointwise distance
\begin{equation}
d^2_H(\Phi,\Psi)(\varrho) := \inf_{\Gamma,\Upsilon \in \cB(\cH)}
\left\{ c(\Gamma-\Upsilon;\rho) : \Gamma^\dag \Gamma = \Phi_\tau,
\Upsilon^\dag \Upsilon = \Psi_\tau \right\}
\label{qqmmdistpt}
\end{equation}
between $\Phi$ and $\Psi$ on the set $\cS(\fg)$ of all normal states on $\cA = \cB(\fg)$. Just
as with the probability kernels in the commutative setting described in the preceding section, $d^2_H(\Phi,\Psi)$ coincides with the supremum of
$d^2_H(\Phi,\Psi)(\varrho)$ over all normal states $\varrho \in
\cS(\fg)$ whenever $\Phi$ and $\Psi$ are (proportional to)
quantum
channels:

\begin{theorem}\label{thm:mmfid}
Let $\map{\Phi,\Psi}{\cB}{\cA}$ be quantum operations with the
respective operational densities $\Phi_\tau,\Psi_\tau \in
\cB(\cH)$. Suppose that for all $\varrho \in \cS(\fg)$ the pairings
\begin{equation}
(\Phi_\tau, \rho \tp \idty_\fh) \equiv \varrho[\Phi(\idty_\fh)], \;
(\Psi_\tau, \rho \tp \idty_\fh) \equiv \varrho[\Psi(\idty_\fh)]
\label{eq:constpair}
\end{equation}
are constant. Then
\begin{equation}
d^2_H(\Phi,\Psi) = \sup_{\varrho \in \cS(\fg)}d^2_H(\Phi,\Psi)(\varrho).
\label{eq:mmfid1}
\end{equation}
Furthermore, then we have that
\begin{equation}
d^2_H(\Phi,\Psi) + \sqrt{\norm{\Phi}\norm{\Psi}} f(\Phi,\Psi) =
\frac{1}{2}(\norm{\Phi} + \norm{\Psi}),
\label{eq:mmfid2}
\end{equation}
where
\begin{equation}
f(\Phi,\Psi) = \inf_{\varrho \in \cS(\fg)} \sup_{\Upsilon \in
\cB(\cH): \Upsilon^\dag \Upsilon = \Psi_\tau} \frac
{
\Re \Tr\big[\Phi_\tau^{1/2}\Upsilon (\rho\ \otimes \idty_\fh)\big]
}
{\sqrt{\varrho\left[\Phi\left(\openone_{\fh}\right)\right]}
\sqrt{\varrho\left[  \Psi\left(  \openone_{\fh}\right)  \right]  }
}
\label{qqofid}
\end{equation}
is the {\em minimax fidelity} between $\Phi$ and $\Psi$.
\end{theorem}

\begin{proof} Fix an arbitrary $\varrho \in \cS(\fg)$. From Eq.~(\ref{eq:constpair}) it follows that
$$
\norm{\Phi} = \sup_{\varrho \in \cS(\fg)} \varrho[\Phi(\idty_\fh)] =
\varrho[\Phi(\idty_\fh)],
$$
and the same goes for $\Psi$. Therefore, given any pair $\Gamma,\Upsilon \in \cB(\cH)$ such that
$\Gamma^\dag \Gamma = \Phi_\tau$ and $\Upsilon^\dag \Upsilon =
\Psi_\tau$, we can write
\begin{eqnarray*}
c(\Gamma-\Upsilon; \rho) &=& \frac{1}{2} \Tr \big((\Gamma -
\Upsilon)^\dag (\Gamma - \Upsilon) (\rho \tp \idty_\fh)\big) \\
&=& \frac{1}{2} \Tr \big( (\Phi_\tau + \Psi_\tau)(\rho \tp
\idty_\fh) - (\Gamma^\dag \Upsilon + \Gamma \Upsilon^\dag)(\rho \tp
\idty_\fh) \big) \\
&=& \frac{1}{2} \Big( \norm{\Phi} + \norm{\Psi} - \Tr[
(\Gamma^\dag \Upsilon + \Gamma \Upsilon^\dag)(\rho \tp \idty_\fh)]\Big),
\end{eqnarray*}
whence it follows that
\begin{eqnarray*}
d^2_H(\Phi,\Psi)(\varrho) &=& \inf_{\substack{\Gamma: \Gamma^\dag \Gamma = \Phi_\tau\\
\Upsilon: \Upsilon^\dag \Upsilon = \Psi_\tau}} c(\Gamma-\Upsilon;
\rho) \\
&=& \frac{1}{2}  \Big( \norm{\Phi} + \norm{\Psi} -
\sup_{\substack{\Gamma: \Gamma^\dag \Gamma = \Phi_\tau\\
\Upsilon: \Upsilon^\dag \Upsilon = \Psi_\tau}}\Tr [(\Gamma^\dag \Upsilon + \Gamma \Upsilon^\dag)(\rho \tp
\idty_\fh)]\Big).
\end{eqnarray*}
Taking the supremum of both sides over all $\varrho \in \cS(\fg)$, we
obtain
\begin{eqnarray}
&& \sup_{\varrho \in \cS(\fg)} d^2_H(\Phi,\Psi)(\varrho) = \frac{1}{2}
\Big( \norm{\Phi} + \norm{\Psi} \nonumber\\
&& \qquad \qquad - \inf_{\varrho \in
  \cS(\fg)}\sup_{\substack{\Gamma: \Gamma^\dag \Gamma =
    \Phi_\tau \\
\Upsilon: \Upsilon^\dag \Upsilon = \Psi_\tau}}\Tr[
(\Gamma^\dag \Upsilon + \Gamma \Upsilon^\dag)(\rho \tp
\idty_\fh)]\Big).
\label{eq:mm1}
\end{eqnarray}
On the other hand,
\begin{eqnarray*}
d^2_H(\Phi,\Psi) &=& \inf_{\substack{\Gamma: \Gamma^\dag \Gamma =
    \Phi_\tau\\
\Upsilon: \Upsilon^\dag \Upsilon = \Psi_\tau}} \sup_{\varrho \in
    \cS(\fg)} c(\Gamma-\Upsilon;\rho) \\
&=& \frac{1}{2} \inf_{\substack{\Gamma: \Gamma^\dag \Gamma =
    \Phi_\tau\\
\Upsilon: \Upsilon^\dag \Upsilon = \Psi_\tau}} \sup_{\varrho \in
    \cS(\fg)} \Big( \norm{\Phi} + \norm{\Psi} - \Tr[
(\Gamma^\dag \Upsilon + \Gamma \Upsilon^\dag)(\rho \tp \idty_\fh)]\Big),
\end{eqnarray*}
which yields
\begin{equation}
d^2_H(\Phi,\Psi) = \frac{1}{2} \Big( \norm{\Phi} + \norm{\Psi} - \sup_{\substack{\Gamma: \Gamma^\dag \Gamma =
    \Phi_\tau\\
\Upsilon: \Upsilon^\dag \Upsilon = \Psi_\tau}} \inf_{\varrho \in
    \cS(\fg)}\Tr[
(\Gamma^\dag \Upsilon + \Gamma \Upsilon^\dag)(\rho \tp
    \idty_\fh)]\Big).
\label{eq:mm2}
\end{equation}
Note that the right-hand sides of Eqs.~(\ref{eq:mm1}) and
(\ref{eq:mm2}) differ only in the order of the
extrema. Thus, establishing the validity of Eq.~(\ref{eq:mmfid1})
amounts to justifying the interchange of the extrema.

According to Lemma~\ref{lm:lagrange}, the supremum over $\Gamma$ and
$\Upsilon$ in
Eq.~(\ref{eq:mm1}) can be evaluated by fixing $\Gamma =
\Phi^{1/2}_\tau$ first and then varying only over all $\Upsilon \in \cB(\cH)$
such that $\Upsilon^\dag \Upsilon = \Psi_\tau$. By the polar
decomposition, any such $\Upsilon$ has the form $U\Psi_\tau^{1/2}$ for
some partial isometry $U$. Thus we have
\begin{eqnarray}
&& \sup_{\substack{\Gamma: \Gamma^\dag \Gamma = \Phi_\tau \\
\Upsilon: \Upsilon^\dag \Upsilon = \Psi_\tau}}\Tr[(\Gamma^\dag \Upsilon + \Gamma \Upsilon^\dag)(\rho \tp
\idty_\fh)] = 2 \sup_{\Upsilon: \Upsilon^\dag \Upsilon = \Psi_\tau} \Re
\Tr [\Phi^{1/2}_\tau\Upsilon (\rho \tp \idty_\fh)] \nonumber \\
&& \qquad \qquad = 2 \sup_U \Re \Tr [\Phi^{1/2}_\tau U\Psi^{1/2}_\tau(\rho
\tp \idty_\fh)],
\label{eq:mm2a}
\end{eqnarray}
where the supremum in Eq.~(\ref{eq:mm2a}) is taken over all partial
isometries $U$ such that
$$
\Psi^{1/2}_\tau U^\dag U \Psi^{1/2}_\tau = \Psi_\tau.
$$
Since the expression being minimized
is linear in $U$ and since any partial isometry can be expressed as a
convex combination of at most four unitaries, we may instead take the supremum
over the unitary group ${\sf U}(\cH)$ and, in fact, over the entire unit ball
${\sf B}_1(\cH) := \{ X \in \cB(\cH) : \norm{X} \le 1\} $:
\begin{equation}
\sup_{\Upsilon: \Upsilon^\dag \Upsilon = \Psi_\tau} \Re
\Tr [\Phi^{1/2}_\tau\Upsilon (\rho \tp \idty_\fh)] = \sup_{X
  \in {\sf B}_1(\cH)}\Re \Tr [\Phi^{1/2}_\tau X\Psi^{1/2}_\tau(\rho
\tp \idty_\fh)].
\label{eq:mm3}
\end{equation}
Since the expression being maximized in the right-hand side of
Eq.~(\ref{eq:mm3}) is affine in both $X$ and $\rho$, and since ${\sf B}_1(\cH)$ and $\cS(\fg)$ are closed convex subsets of $\cB(\cH)$ and
$\cB_\intercal(\fg)$ respectively, it follows from standard minimax
arguments \cite{Lue69} that we can indeed interchange the
extrema to obtain $f_-(\Phi,\Psi) = f_+ (\Phi,\Psi)$, where
\begin{eqnarray*}
f_-(\Phi,\Psi) &:=& \inf_{\varrho \in \cS(\fg)} \sup_{X \in {\sf B}_1(\cH)}\Re \Tr [\Phi^{1/2}_\tau X\Psi^{1/2}_\tau(\rho
\tp \idty_\fh)] \\
f_+(\Phi,\Psi) &:=& \sup_{X \in {\sf B}_1(\cH)}\inf_{\varrho \in \cS(\fg)} \Re \Tr [\Phi^{1/2}_\tau X\Psi^{1/2}_\tau(\rho
\tp \idty_\fh)],
\end{eqnarray*}
which proves the claim of Eq.~(\ref{eq:mmfid1}). The rest is
straightforward. \end{proof}
As seen immediately from Theorem~\ref{thm:mmfid}, when $\Phi$ and
$\Psi$ are quantum channels, then
$$
d^2_H( \Phi,\Psi )  +f(  \Phi,\Psi)  =1,
$$
with the minimax fidelity given by
\begin{equation}
f\left(  \Phi,\Psi\right)  =\inf_{\varrho\in\mathcal{S}\left(  \fg%
\right)  }\sup_{\Upsilon:\Upsilon^{\dagger}\Upsilon=\Psi_\tau%
}\Re \Tr\big[  \Phi^{1/2}_\tau\Upsilon\left(
\rho\otimes \openone_{\fh}\right)  \big]  .
\label{qqmmfid}
\end{equation}

\section{Evaluating the fidelity distances }
\label{sec:evaluate}

\subsection{Fidelities for quantum states and quantum effects}
\label{ssec:qeff}

Consider two normal states $\varrho,\varsigma$ on $\cB = \cB(\fh)$ as quantum channels from $\cB$
into the Abelian algebra $\cA = \cB(\fg)$ with $\fg \simeq \C$. In
this case, the operational
densities $\varrho_\tau$, $\varsigma_\tau$ of $\varrho, \varsigma$
coincide with the corresponding density
operators $\rho$, $\sigma$: $\varrho_\tau = \rho$
and $\varsigma_\tau = \sigma$. The predual maps
$\map{\varrho_\intercal,\varsigma_\intercal}{\cA_\intercal \simeq \C}{\cB_\intercal}$
can then be thought of as the {\em state creation operations},
$\varrho_\intercal(\lambda) = \lambda \rho$ and
$\varsigma_\intercal(\lambda) = \lambda \sigma$ for $\lambda \in \C$.

 In order to compute the minimax fidelity
$f(\varrho,\varsigma)$, we have to consider all $\chi,\psi
\in \cB$ that give the decompositions $\rho = \chi^\dag \chi$ and $\sigma = \psi^\dag
\psi$.  Note that we can always write these decompositions as purifications
$$
\rho=\sum_{j}|\chi_{j}\rangle\langle\chi_{j}|,\;\sigma=\sum_{j}|\psi_{j}\rangle\langle\psi_{j}|,
$$
where $\ket{\chi_j} := \chi\ket{j}$, $\ket{\psi_j} := \psi\ket{j}$ with respect to a fixed
orthonormal basis $\left\{ \ket{j}\right \}$ of $\fh$. We then have the minimum
quadratic distance
\begin{eqnarray*}
d^2_H(\varrho,\varsigma) &=& \frac{1}{2}\inf_{\substack{\chi \in \cB:
    \chi^\dag\chi=\rho\\
\psi \in \cB: \psi^\dag\psi=\sigma}}\sup_{\varpi \in
    \cS(\fg)}\varpi\big[(\chi-\psi)^\dag(\chi-\psi)\big] \\
&\equiv&
    \frac{1}{2} \inf_{\substack{\chi \in \cB: \chi^\dag \chi = \rho\\
\psi \in \cB: \psi^\dag\psi = \sigma}} \Tr \big[(\chi - \psi)^\dag (\chi -
    \psi)\big],
\end{eqnarray*}
where the last equality is due to the fact that $\dim \fg =
1$. Expanding the product under the trace, we can write
\begin{eqnarray}
d^2_H(\varrho,\varsigma) &=& \frac{1}{2}\left[\Tr (\rho + \sigma) -
  \sup_{\chi,\psi \in \cB}\left\{\Re \Tr (\chi^\dag
  \psi) : \chi^\dag \chi = \rho, \psi^\dag \psi =
  \sigma \right\} \right] \label{eq:stdist1} \\
&=& 1 -  \sup_{\substack{\chi \in \cB: \chi^\dag \chi = \rho\\
\psi \in \cB: \psi^\dag \psi = \sigma}}\Re \Tr (\chi^\dag
  \psi) \label{eq:stdist2} \\
&\equiv& 1 - f(\varrho,\varsigma).\label{eq:stdist3}
\end{eqnarray}
According to Lemma~\ref{lm:lagrange}, the supremum in Eq.~(\ref{eq:stdist2}) is attained at any $\chi\in\cB$ satisfying the condition $\chi^\dag \chi
= \rho$, say $\chi
=\rho^{1/2}$, and $\psi=\psi_\opt $ satisfying the equation
$\psi_\opt \chi^\dag=(\chi\sigma\chi^\dag)^{1/2}=\chi
\psi_\opt ^\dag$:
\begin{eqnarray*}
f(\varrho,\varsigma) &=& \sup_{\substack{\chi \in \cB: \chi^\dag \chi = \rho\\
\psi \in \cB: \psi^\dag \psi = \sigma}}\Re \Tr( \chi^\dag
  \psi) \\
&=& \sup_{\psi\in\cB(\fh)}
\left\{  \Re \Tr( \rho^{1/2}\psi)
:\psi^{\dagger}\psi=\sigma\right\} \\
&=& \Tr \sqrt{\rho \sigma}.
\end{eqnarray*}
Observe that the standard Uhlmann fidelity between
the density operators $\rho$ and $\sigma$, $F(\rho,\sigma)$ in Eq.~(\ref{eq:statefid}), can be written as $F(\rho,\sigma) =
\trnorm{\rho^{1/2} \sigma^{1/2}} = \Tr \sqrt{\rho \sigma}$.
Thus the minimax fidelity between two normal states
$\varrho$ and $\varsigma$ on $\cB$, or, equivalently, between the state
creation operations
$\map{\varrho_\intercal,\varsigma_\intercal}{\C}{\cB_\intercal(\fh)}$, agrees
with the Uhlmann fidelity between the respective density operators
$\rho$ and $\sigma$ of $\varrho$ and $\varsigma$.

Next we turn to the other extreme case, namely that of the {\em state
  annihilation operations} $\Phi,\Psi$ with the preduals
  $\Phi_\intercal(\rho) =(\Phi_\tau,\rho)$, $\Psi_\intercal(\rho)
  =(\Psi_\tau,\rho)$, corresponding to $\dim\fh=1$. They are
  completely specified by the {\em effects}, i.e., the positive operators
  $\Phi_\tau,\Psi_\tau\in\cB(\fg)$ satisfying $0 \le
  \Phi_\tau,\Psi_\tau \le \idty_\fg$, which can be purified as in (\ref{qqpurif}), where $\Gamma_{j}=\langle
j|\Gamma$, $\Upsilon_{j}=\langle j|\Upsilon$ are the bra-vectors corresponding
to an othonormal basis $\{ \ket{j} \}$ in $\fg$. The squared pointwise
  minimax
  distance between the state annihilation operations $\Phi,\Psi$, or,
  equivalently between the effects $\Phi_\tau,\Psi_\tau$, on the set
  $\cS(\fg)$ of normal states $\varrho = \rho^\intercal$ on $\cB(\fg)$
  is given by the minimum
$$
d^2_H(\Phi,\Psi)(\varrho)  =\frac{1}{2}\inf
_{\Gamma,\Upsilon \in \cB(\fg)}
\left\{\Tr\left[(\Gamma-\Upsilon)^\dag(\Gamma-\Upsilon)\rho\right] :
\Gamma^\dag \Gamma = \Phi_\tau, \Upsilon^\tau \Upsilon = \Psi_\tau \right\}
$$
of the quadratic distance between their purifications
$\Gamma,\Upsilon\in\cB(\fg)$. The solution of this problem is likewise
given by Lemma~\ref{lm:lagrange} with $R=\Phi_\tau$ and
$S=\rho\Psi_\tau\rho$. Thus the optimum
$$
d^2_H(\Phi,\Psi)(\varrho) =
\frac{1}{2}\Tr[(\Phi_\tau+\Psi_\tau)\rho]-\Tr\sqrt{\Phi_\tau(\rho\Psi_\tau\rho)}
$$
is attained at any $\Gamma\in\cB$ satisfying the condition
$\Gamma^\dag \Gamma = \Phi_\tau$, say $\Gamma=\Phi_\tau^{1/2}$,
and the corresponding $\Upsilon=\Upsilon_\opt$
satisfying the equation
$\Upsilon_\opt \rho\Gamma^\dag=\sqrt{\Gamma\rho\Psi_\tau\rho\Gamma^\dag}
= \Gamma\rho\Upsilon^\dag_\opt$. The maximum of this distance
over all states,
\begin{eqnarray*}
d^2_H(\Phi,\Psi) &=& \sup_{\varrho \in \cS(\fg)} d^2_H(\Phi,\Psi)(\varrho) \\
&\equiv& \sup_{\varrho \in \cS(\fg)}\left(  \frac{1}{2}\Tr[(\Phi_\tau +
  \Psi_\tau) \rho]-\Tr\sqrt{\Phi_\tau (\rho \Psi_\tau \rho)}\right) \\
&=& \sup_{\varrho \in \cS(\fg)} \inf_{\Gamma,\Upsilon \in \cB(\fg)}
\left\{ \Tr [(\Gamma-\Upsilon)^\dag (\Gamma- \Upsilon) \rho] : \Gamma^\dag \Gamma = \Phi_\tau, \Upsilon^\dag \Upsilon =
\Psi_\tau \right\},
\end{eqnarray*}
is given by the minimax quadratic distance
$$
d^2_H(\Phi,\Psi) = \frac{1}{2}\inf_{\Gamma,\Upsilon \in
  \cB(\fg)}\left\{ \left\|\Gamma-\Upsilon \right\|
^2 : \Gamma^\dag \Gamma = \Phi_\tau, \Upsilon^\dag \Upsilon =
\Psi_\tau \right\},
$$
interchange of the extrema following from standard minimax arguments
\cite{Lue69} and the fact that all $\Gamma,\Upsilon$ satisfying,
respectively, $\Gamma^\dag \Gamma = \Phi_\tau$ and $\Upsilon^\dag
\Upsilon = \Psi_\tau$ are contained in the unit ball of $\cB(\fg)$.

\subsection{Semiclassical fidelity}
\label{ssec:semiclass}

It is straightforward to extend the formalism of Section~\ref{ssec:kernels} involving the
commutative Hellinger distance between two positive kernels to the case of mappings from a set $X$
into positive trace-class operators on the Hilbert space $\fh$, i.e.,
$\boldsymbol{\rho}:{x \in X}\longmapsto \rho(x) \in \cB_\intercal(\fh)$
and $\boldsymbol{\sigma}:{x \in X}\longmapsto \sigma(x) \in
\cB_\intercal(\fh)$ with $\rho(x),\sigma(x) \ge
0$ for all $x \in X$. We thus have the pointwise Hellinger distance 
$$
d^2_H( \rho(x),\sigma(x))  =\left(\openone,\frac{1}{2}[\rho(x) + \sigma(x)]\right)
- \sqrt{(\openone,\rho(x)) (\openone,\sigma(x))}f(\rho(x),\sigma(x)) 
$$
in terms of the trace pairing $( B,\rho ) =\Tr (B\tran{\rho})$ of
$B \in \cB=\cB(\fh)$ and $\cB_\intercal=\cB_\intercal(\fh)$, where
$$
f(\rho(x), \sigma(x))
= \frac{( \openone,\sqrt{\rho(x)
    \sigma(x)})} {\sqrt{(\openone,\rho(x))(\openone,\sigma(x))}}
 =\frac{\Tr\sqrt{\rho(x)\sigma(x)}}{\sqrt{\Tr\rho(x)\Tr\sigma(x)}}.
$$
The semi-classical operational distance between
$\boldsymbol{\rho}=\rho(\bullet)$ and
$\boldsymbol{\sigma}=\sigma(\bullet)$ can then be defined as
\begin{equation}
d^2_H( \boldsymbol{\rho},\boldsymbol{\sigma}) = \sup_{x \in X} d(\rho(x),
\sigma(x)) \equiv \| d(\rho(\bullet),\sigma(\bullet)) \|.
\label{cqdist}
\end{equation}
When $\Tr \rho(x) = 1 = \Tr \sigma(x)$ for all $x \in X$, i.e.,
when $\boldsymbol{\rho}$ and $\boldsymbol{\sigma}$ are {\em
  classical-to-quantum}, c-q (or {\em semiclassical}) channels,
Eq.~(\ref{cqdist}) can be written as
$d^2_H(\boldsymbol{\rho},\boldsymbol{\sigma}) = 1 -
f(\boldsymbol{\rho},\boldsymbol{\sigma})$, where
\begin{eqnarray*}
f(\boldsymbol{\rho},\boldsymbol{\sigma}) &=& \inf_{x \in X}
\Tr[\rho(x)^{1/2}\sigma(x)\rho(x)^{1/2}]^{1/2}\\
&=& \inf_{x
  \in X}\Tr\sqrt{\rho(
x)\sigma(x)} \\
&\equiv& \inf_{x \in X}F(\rho(x),\sigma(x))
\label{cqcfid}
\end{eqnarray*}
is the minimax fidelity of $\boldsymbol{\sigma}$ relative to
$\boldsymbol{\rho}$.

\subsection{Semiquantum fidelity}
\label{ssec:semiquant}

Next we consider the opposite of semiclassical operations --- namely,
the {\em semiquantum operations} which correspond to quantum
measurements as quantum-to-classical (q-c) channels. Such operations
are given as
$$
\Phi(b) = \int_Y b(y) \Phi_\tau(y) \mathrm{d}\mu(y)
\equiv( b,\Phi_\tau)
$$
on the algebra $\cB = \mathcal{C}(Y)$ of continuous bounded functions
$\map{b}{Y}{\C}$, where $(Y,\cB_Y,\mu)$ is a measure space, by specifying the positive operator-valued Bochner $\mu$-integrable functions $\map{\Phi_\tau}{Y}{\cA=\cB(\fg)}$. If
$$
\Phi(1)  =(1,\Phi_\tau) =\openone_{\fg},
$$
the predual maps $\cA \ni \rho \longmapsto \Phi_\intercal (\rho)(\bullet)  \in\mathcal{C}_\intercal(Y)$,
$$
\Phi_\intercal(\rho)(y)  :=\big(\Phi_\tau(y),\rho \big)  \equiv
\varrho\left[  \Phi_\tau(y)  \right],
$$
define for each input quantum state $\varrho \in \cS(\fg)$ a classical
probability density on $(Y,\cB_Y,\mu)$, that is, they describe quantum
measurements by the positive operator-valued measures (POVM's) $M(\mathrm{d}y)
= \Phi_\tau(y) \mathrm{d}\mu(y)$.

In order to avoid technicalities in defining the semi-quantum fidelity
distance between two q-c channels $\map{\Phi,\Psi}{\cB}{\cA}$, we shall assume that $\Phi_\tau(y),\Psi_\tau(y)$ are weakly continuous bounded functions
on $Y$. Then the squared distance $d^2_H(\Phi,\Psi)$ can be written as
\begin{equation}
d^2_H(\Phi,\Psi) = \inf_{\Gamma, \Upsilon: \Gamma^\dag\Gamma=\Phi_\tau,
  \Upsilon^\dag \Upsilon =\Psi_\tau}
\left\| \int\big( \Gamma(y) - \Upsilon(y)\big)^\dag
\big( \Gamma(y) - \Upsilon(y)  \big)  \mathrm{d}\mu(y)  \right\|,
\label{qcfdist}
\end{equation}
where the decompositions $\Gamma^\dag \Gamma = \Phi_\tau$ and
$\Upsilon^\dag \Upsilon = \Psi_\tau$ are understood in the pointwise
sense as
$$
\Phi_\tau(y) = \Gamma(y)^\dag\Gamma(y), \Psi_\tau(y) =
\Upsilon(y)^\dag\Upsilon(y), \qquad \forall y \in Y.
$$
The infimum in Eq.~(\ref{qcfdist}) is achieved at any $\Gamma \in \cA \otimes \mathcal{C}_\intercal(Y)$
satisfying the condition $\Gamma^\dag \Gamma = \Phi_\tau$, say
$\Gamma(y)=\Phi_\tau(y)^{1/2}$, and the corresponding $\Upsilon=\Upsilon
_\opt $ satisfying the equation
$$
\Upsilon_\opt (y) \rho \Gamma(y)^\dag = 
[\Gamma(y)\rho\Psi_\tau(y)\rho\Gamma(y)^\dag]^{1/2}
= \Gamma(y)\rho\Upsilon_\opt (y)^\dag.
$$
The maximum of this minimal distance over all states,
$$
d^2_H(\Phi,\Psi) = \sup_{\varrho \in \cS(\fg)}
\int\left(  \frac{1}{2}\Tr\left[\big(\Phi_\tau(y) + \Psi_\tau(y)\big)\rho\right]-\Tr
\sqrt{\Phi_\tau(y)\big(\rho\Psi_\tau(y)\rho\big)}\right)  \mathrm{d}\mu\left(  y\right)  ,
$$
is equal to $d^2_H(\Phi,\Psi) = 1-f(\Phi,\Psi)$ in the
measurement operation case $\Phi(1) = \openone_{\fg} = \Psi(1)$, where
\begin{equation}
f(\Phi,\Psi)  =\inf_{\varrho\in\cS(\fg)}
\int\Tr\sqrt{\Phi_\tau(y)\big(\rho\Psi_\tau(y)\rho\big)}\mathrm{d}\mu(y).
\label{qccfid}
\end{equation}

\subsection{Operational fidelity formula}
\label{ssec:opfidformula}

Now we can easily evaluate the minimax formula (\ref{qqmmdist}) for the
fidelity of two general quantum operations
$\map{\Phi,\Psi}{\cB}{\cA}$, $\cB = \cB(\fh)$, $\cA = \cB(\fg)$. The
solution of this problem is also given by Lemma~\ref{lm:lagrange} with
$R = \Phi_\tau$ and $S=(\rho \otimes \openone_\fh) \Psi_\tau
(\rho \otimes \openone_\fh)$. For a given $\varrho \in \cS(\fg)$, the
supremum in
\begin{eqnarray*}
&& d^2_H(\Phi,\Psi)(\varrho) =\frac{1}{2}\Big(\Tr\big[(\Phi_\tau +
  \Psi_\tau)(\rho\otimes \idty_\fh) \big] \\
&& \qquad \qquad - 2 \sup_{\Gamma,\Upsilon
    \in \cB(\cH)} \left\{\Re \Tr \big[\Gamma^\dag \Upsilon (\rho \tp
  \idty_\fh)\big] : \Gamma^\dag \Gamma = \Phi_\tau, \Upsilon^\dag \Upsilon =
  \Psi_\tau \right\}\Big)
\end{eqnarray*}
is equal to $\Tr \sqrt{\Phi_\tau \left[(\rho \tp \idty_\fh)
  \Psi_\tau (\rho \tp \idty_\fh)\right] }$, and
is achieved at any $\Gamma\in \cB(\cH)$ satisfying the
condition $\Gamma^\dag \Gamma = \Phi_\tau$, say
$\Gamma=\Phi^{1/2}_\tau$, and the corresponding $\Upsilon=\Upsilon_\opt$ satisfying the equation
$$
\Upsilon_\opt (\rho \otimes \openone_\fh) \Gamma^\dag =
[\Gamma(\rho \otimes \idty_\fh) \Psi_\tau (\rho \otimes \idty_\fh)
\Gamma^\dag ]^{1/2} =\Gamma (\rho \otimes \idty_\fh) \Upsilon^\dag_\opt.
$$
When $\Phi,\Psi$ are quantum channels, or, equivalently, when the
preduals $\Phi_\intercal,\Psi_\intercal$ are trace-preserving,
Theorem~\ref{thm:mmfid} says that the maximum of this distance over all states,
\begin{equation}
d^2_H(\Phi,\Psi) = \sup_{\varrho\in\cS(\fg)}
\Tr\left( \frac{1}{2}(\Phi_\tau + \Psi_\tau)(\rho \otimes \idty_\fh) -
  \sqrt{\Phi_\tau[(\rho \otimes \idty_\fh) \Psi_\tau (\rho \otimes
    \idty_\fh)] }\right),
\label{qqfdistev}
\end{equation}
can be written as $d^2_H(\Phi,\Psi) = 1-f(\Phi,\Psi)$, where
\begin{equation}
f(\Phi,\Psi) = \inf_{\varrho \in \cS(\fg)}\Tr\sqrt{\Phi_\tau [
  (\rho \otimes \idty_\fh) \Psi_\tau (\rho \otimes \idty_\fh)\big]}
\label{qqcfid}
\end{equation}
is the minimax fidelity between $\Phi$ and $\Psi$.

\subsection{Operational fidelity in terms of Kraus and Stinespring decompositions}
\label{ssec:qofid_kraus}

Consider, as before, two quantum channels $\map{\Phi,\Psi}{\cB}{\cA}$,
where $\cB = \cB(\fh)$ and $\cA = \cB(\fg)$. Given the minimax fidelity
\begin{eqnarray*}
f(\Phi,\Psi) &=& \inf_{\varrho \in \cS(\fg)} \sup_{
\substack{\Gamma: \Gamma^\dag
\Gamma = \Phi_\tau\\
\Upsilon: \Upsilon^\dag \Upsilon = \Psi_\tau} } \Re \Tr
[\Gamma^\dag \Upsilon (\rho \tp
\idty_\fh)] \\
&=& \inf_{\varrho \in \cS(\fg)} \sup_{\substack{\Gamma:
\Gamma^\dag \Gamma = \Phi_\tau\\
\Upsilon: \Upsilon^\dag \Upsilon = \Psi_\tau}} \Abs{\Tr[\Gamma^\dag \Upsilon (\rho \tp \idty_\fh)]}
\end{eqnarray*}
between $\Phi$ and $\Psi$, where the second equality follows from
Lemma~\ref{lm:homf}, the supremum over all $\Gamma$ and
$\Upsilon$ satisfying, respectively, $\Gamma^\dag \Gamma = \Phi_\tau$
and $\Upsilon^\dag \Upsilon = \Psi_\tau$ can be replaced with the
supremum over all Kraus decompositions of $\Phi$ and $\Psi$, i.e.,
over all collections $\{ F_j \}$, $\{V_j\}$ of bounded operators $\fg
\longrightarrow \fh$, determined from $\Phi_\tau,\Psi_\tau$ via
Eqs.~(\ref{qqpurif}) and (\ref{eq:kraus}):
\begin{equation}
f(\Phi,\Psi) = \inf_{\varrho \in \mathcal{S}(\fg)}
\sup_{\{F_j\},\{V_j\}} \big| \sum_j\varrho(F^\dag_j
  V_j)\big|.
\label{eq:qfidkraus}
\end{equation}
Just as in the proof of Theorem~\ref{thm:mmfid}, we may
restrict ourselves only to those $\Gamma,\Upsilon$ that can be written as
$\Gamma = U\Phi^{1/2}_\tau, \Upsilon = V\Psi^{1/2}_\tau$ for some
unitaries $U,V$. Thus, if we write
$\Phi^{1/2}_\tau$ and $\Psi^{1/2}_\tau$ in the form of
Eq.~(\ref{eq:kraus}) as
$$
\Phi^{1/2}_\tau = \sum_j \ket{j}(\hat{F}_j|, \quad \Psi^{1/2}_\tau = \sum_j
\ket{j}(\hat{V}_j|,
$$
then it follows that, given a unitary $U$, we can write
$$
\Gamma = U\Phi^{1/2}_\tau = \sum_j \ket{j} \big(\sum_\ell
U_{j\ell} \hat{F}_\ell \big| \equiv \sum_j \ket{j} (\hat{F}_j(U)|,
$$
and similarly for $\Upsilon = V\Psi^{1/2}_\tau$. Thus
\begin{eqnarray*}
f(\Phi,\Psi) &=& \inf_{\varrho \in \cS(\fg)}
\sup_{U,V \in {\sf U}(\cH)} \Big| \sum_j\varrho
[\hat{F}_j(U)^\dag \hat{V}_j(V)]\Big| \\
&=& \inf_{\varrho \in
\cS(\fg)} \sup_{U \in {\sf U}(\cH)} \Big| \sum_j\varrho[\hat{F}_j(U)^\dag \hat{V}_j]\Big|.
\end{eqnarray*}
Turning now to the infimum over all normal states $\varrho$ on
$\cA \equiv \cB(\fg)$, we may equivalently consider all pairs
$\{\varphi,\mathcal{K}\}$, where $\varphi$ is a normal $*$-representation of
$\cA$ on a Hilbert space $\cK$:
$$
f(\Phi,\Psi) := \inf_{\{\varphi,\mathcal{K}\}; \upsilon \in
\mathcal{K}, \| \upsilon \| = 1}
\sup_{U \in {\sf U}(\cH)} \Big\vert \sum_j \Braket{\upsilon}{\varphi[\hat{F}_j(U)^\dag \hat{V}_j]}{\upsilon} \Big\vert.
$$
Since all normal $*$-representations of the full operator algebra
$\cB(\fg)$ are unitarily equivalent to an amplification
$B \longmapsto B \otimes \openone_{\fk}$ for some Hilbert space
$\fk$, we can write
\begin{equation}
f(\Phi,\Psi) := \inf_{\upsilon \in \fg \otimes \fk;
\| \upsilon \| = 1}
\sup_{U \in {\sf U}(\cH)} \Big\vert \sum_j
\Braket{\upsilon}{\hat{F}_j(U)^\dag \hat{V}_j
\otimes \openone_{\fk}}{\upsilon}\Big\vert.
\label{eq:minimax2a}
\end{equation}
Introducing the vectors $|\upsilon,\Phi\rangle, |\upsilon,\Psi\rangle \in
\fg \otimes \fk \otimes \mathcal{H}$, defined by
\[
| \upsilon,\Phi \rangle := \sum_j (\hat{F}_j \otimes
  \openone_\fk)\upsilon \otimes |j\rangle, \quad |\upsilon,\Psi
  \rangle := \sum_j (\hat{V}_j \otimes \openone_\fk)\upsilon
  \otimes |j \rangle,
\]
we obtain yet another form of the minimax fidelity:
\begin{equation}
f(\Phi,\Psi) =  \inf_{\upsilon \in \fg \otimes \fk}
\sup_{U \in {\sf U}(\cH)} \vert \langle \upsilon, \Phi | \openone_{\fg
\otimes \fk} \otimes U | \upsilon, \Psi \rangle \vert.
\end{equation}
For a fixed $\upsilon \in \fg \otimes \fk$, taking the
supremum over $U$ is tantamount to taking the supremum of $|\langle
\chi | \xi \rangle |$ over all pairs of unit vectors $\chi,\xi \in \fg
\otimes\fk \otimes \mathcal{H}$ such that
\begin{eqnarray*}
\Tr_{\mathcal{H}} |\chi \rangle \langle \chi | &=& \sum_j
\big(\tran{\hat{F}_j} \otimes
\openone_{\fk}\big)|\upsilon \rangle \langle \upsilon |
\big(\tran{\hat{F}_j} \otimes \openone_{\fk}\big)^\dag
\equiv \Phi_\intercal \otimes \operatorname{id} (| \upsilon \rangle
\langle \upsilon |), \\
\Tr_{\mathcal{H}} |\xi \rangle \langle \xi | &=& \sum_j
\big( \tran{\hat{V}_j} \otimes
\openone_{\fk}\big)|\upsilon \rangle \langle \upsilon |
\big(\tran{\hat{V}_j} \otimes \openone_{\fk}\big)^\dag
\equiv \Psi_\intercal \otimes \operatorname{id} (| \upsilon \rangle
\langle \upsilon |),
\end{eqnarray*}
which, in conjunction with the standard results on the Uhlmann fidelity
(\ref{eq:statefid}) between density operators \cite{Joz94,Uhl76}, finally yields
\begin{eqnarray*}
f(\Phi,\Psi) &=& \inf_{\upsilon \in \fg \otimes \fk :
  \| \upsilon \| = 1} F\big(\Phi_\intercal \otimes \id
  (\ketbra{\upsilon}{\upsilon}), \Psi_\intercal \otimes \id (\ketbra{\upsilon}{\upsilon})\big) \\
&=& \inf_{\varrho \in \cS(\fg \tp \fk)} F\big(\Phi_\intercal
  \tp \id (\rho), \Psi_\intercal \tp
  \id (\rho)\big).
\end{eqnarray*}
Note that we may always take $\fk$ isomorphic to $\fg$:
\begin{equation}
f(\Phi,\Psi) = \inf_{\upsilon \in \fg \tp \fg, \norm{\upsilon} = 1} F\big(\Phi_\intercal \otimes \id
  (\ketbra{\upsilon}{\upsilon}), \Psi_\intercal \otimes \id
  (\ketbra{\upsilon}{\upsilon})\big).
\label{eq:qofid_main}
\end{equation}

Given some Kraus decompositions $\{ F_j \}$, $\{ V_j \}$ of $\Phi$ and
$\Psi$ respectively, we may define the operators
$$
F\xi := \sum_j F_j \xi \tp \ket{j},\; V\xi := \sum_j V_j \xi \tp
\ket{j}
$$
from $\fg$ into $\fh \tp \cH$ and express $\Phi$ and $\Psi$ in the
Stinespring form $\Phi(B) = F^\dag (B \tp \idty_\cH)F$, $\Psi(B) =
V^\dag (B \tp \idty_\cH)V$ (cf.~Section~\ref{ssec:opfid}). Then we may
rewrite Eq.~(\ref{eq:qfidkraus}) as
$$
f(\Phi,\Psi) = \inf_{\varrho \in \cS(\fg)} \sup_{F,V}
\Abs{\Tr (F\rho V^\dag)},
$$
where the supremum is over all $\map{F,V}{\fg}{\fh\tp\cH}$ giving the Stinespring decompositions of $\Phi$ and $\Psi$ respectively. We
may, as before, fix $F$ and $V$, say, by considering the `canonical'
Kraus decompositions $\{\hat{F}_j\}$, $\{ \hat{V}_j \}$, and instead
take the supremum over all unitaries $U \in {\sf U}(\cH)$:
\begin{eqnarray*}
f(\Phi,\Psi) &=& \inf_{\varrho \in \cS(\fg)} \sup_{U \in {\sf U}(\cH)}
\Abs{\Tr [(\idty_\fh \tp U)F\rho V^\dag]} \nonumber \\
&= & \inf_{\varrho \in \cS(\fg)} \sup_{U \in {\sf U}(\cH)} \Abs{\Tr [U
\Tr_\fh (F\rho V^\dag)]},
\end{eqnarray*}
which yields another useful formula
\begin{equation}
f(\Phi,\Psi) = \inf_{\varrho \in \cS(\fg)} \trnorm{\Tr_\fh (F\rho
V^\dag)}
\label{eq:qofid_stine}
\end{equation}
for the minimax fidelity between the channels $\Phi,\Psi$. It is, in
fact, not hard to show that the right-hand side of
Eq.~(\ref{eq:qofid_stine}) does not depend on the particular choice of
the Stinespring operators $F,V$, as long as we agree to dilate the input
Hilbert space $\fh$ by the `canonical' auxiliary Hilbert space $\cH =
\fg \tp \fh$. 

We note that the constructions of this section are valid more generally for channels
given in terms of the continual Kraus decompositions
$$
\Phi(B) = \int_Z F(z)^\dag B F(z) \mathrm{d}\mu(z),\; \Psi(B) = \int_Z
V(z)^\dag B V(z) \mathrm{d}\nu(z),
$$
provided that the measures $\mu$ and $\nu$ are equivalent, i.e.,
absolutely continuous with respect to each other. Then
Eq.~(\ref{eq:qfidkraus}) is a special instance of the more general expression
$$
f(\Phi,\Psi) = \inf_{\varrho \in \cS(\fg)} \sup_{\{ F(z) \}, \{ V(z)
  \}} \Abs{ \varrho \left(\int_Z \sqrt{\mathrm{d}\nu/\mathrm{d}\mu}
  F(z)^\dag V(z) \mathrm{d}\mu(z) \right)},
$$
where $\mathrm{d}\nu/\mathrm{d}\mu$ is the Radon--Nikodym derivative
of $\nu$ with respect to $\mu$, for the case when both $\mu$ and $\nu$
are counting measures, $\mathrm{d}\mu = \mathrm{d}\nu = 1$, on a
finite or countably infinite set.

\section{Properties of the operational fidelity }
\label{sec:properties}

In this section we establish several key properties of the minimax
fidelity between quantum operations. These properties follow almost
immediately from the corresponding properties enjoyed by the fidelity
(\ref{eq:statefid}) on density operators:
\newcounter{Lcount}
\begin{list}{(F.\arabic{Lcount})}{\usecounter{Lcount}}
\item $F$ is symmetric, $F(\rho,\sigma) = F(\sigma,\rho)$, bounded between 0 and 1, and $F(\rho,\sigma)
= 1$ if and only if $\rho = \sigma$;\label{sym}
\item $F$ is jointly concave over all pairs of density operators;\label{concave}
\item $F$ is unitarily invariant, i.e., $F(\rho,\sigma) =
F(U\rho U^\dag, U\sigma U^\dag)$ for any unitary $U$;\label{uint}
\item $F$ is monotone with respect to quantum channels:
$F\big(\Phi_\intercal(\rho),\Phi_\intercal(\sigma)\big) \ge
F(\rho,\sigma)$ for every quantum channel $\Phi$.\label{monotone}
\item the {\em Bures distance} $d_B(\bullet,\bullet) :=
  \sqrt{1-F(\bullet,\bullet)}$ is topologically equivalent to the
  trace-norm half-distance $D(\bullet,\bullet)$:
  $$
2^{-1/2}D(\rho,\sigma) \le d_B(\rho,\sigma) \le
\sqrt{D(\rho,\sigma)}
$$
[cf.~Eq.~(\ref{eq:fg})].\label{cont}
\end{list}
Property (F.\ref{concave}), in fact, follows from {\em strong concavity} of
$F$ \cite{NC00}, i.e.,
\begin{equation}
F(\sum_i p_i \rho_i , \sum_i q_i \sigma_i ) \ge \sum_i \sqrt{p_i q_i}
F(\rho_i,\sigma_i)
\label{eq:strconcave}
\end{equation}
for all $0 \le p_i,q_i \le 1$ such that $\sum_i p_i = 1 = \sum_i
q_i$.

Using Eq.~(\ref{eq:qofid_main}), we can immediately obtain for the  minimax fidelity $f(\bullet,\bullet)$ on pairs
of quantum channels the following analogues of properties
(F.\ref{sym})--(F.\ref{monotone}) of the fidelity $F(\bullet,\bullet)$ on
pairs of density operators:
\setcounter{Lcount}{0}
\begin{list}{(f.\arabic{Lcount})}{\usecounter{Lcount}}
\item $f$ is symmetric, bounded between 0 and 1, and $f(\Phi,\Psi) =
1$ if and only if $\Phi = \Psi$;
\item $f$ is jointly concave over all pairs of channels;
\item $f$ is invariant under both left and right composition with
unitarily implemented channels, i.e.,
$$
f\big(\Theta_U\circ\Phi,
\Theta_U\circ \Psi\big) = f(\Phi,\Psi)
$$
and
$$
f\big(\Phi \circ \Theta_V,\Psi\circ \Theta_V\big) =
f(\Phi,\Psi)
$$
for any two channels
$\map{\Phi,\Psi}{\cB(\fh)}{\cB(\fg)}$ and any two unitaries $U \in
{\sf U}(\fg)$, $V \in {\sf U}(\fh)$, where $\Theta_U (B) := U^\dag B U$, and
$\Theta_V$ is defined analogously;
\item $f$ is monotone with respect to both left and right composition
with quantum channels, i.e., $f(\Phi \circ \Phi_1, \Psi \circ \Phi_1)
\ge f(\Phi,\Psi)$ and $f(\Phi_2 \circ \Phi, \Phi_2 \circ \Psi) \ge f
(\Phi, \Psi)$ for any two channels $\map{\Phi,\Psi}{\cB}{\cA}$, all
channels $\Phi_1$ into $\cB$, and all channels $\Phi_2$ on $\cA$.
\end{list}
Just as in the case of the fidelity between density operators, the
minimax fidelity $f$ possesses the strong concavity property
\begin{equation}
f\big(\sum_i p_i \Phi_i, \sum_i q_i \Psi_i \big) \ge \sum_i \sqrt{p_i
  q_i} f(\Phi_i,\Psi_i).
\label{eq:strconcave_mm}
\end{equation}
On the other hand, deriving for the minimax fidelity $f$ an analogue of property
(F.\ref{cont}) of the Uhlmann fidelity $F$ requires a bit more
work. To this end, let us consider two channels $\map{\Phi,\Psi}{\cB}{\cA}$, $\cB = \cB(\fh)$,
$\cA = \cB(\fg)$. Suppose first that $\fg$ is infinite-dimensional and
separable. Then $\fg \simeq \ell^2$, and we can rewrite
Eq.~(\ref{eq:qofid_main}) as
$$
f(\Phi,\Psi) = \inf_{\upsilon \in \fg \tp \ell^2; \norm{\upsilon} = 1} F\big(\Phi_\intercal
\tp \id (\ketbra{\upsilon}{\upsilon}), \Psi_\intercal \tp \id
(\ketbra{\upsilon}{\upsilon}) \big).
$$
The space $\ell^2$ contains, as a dense subset, the pre-Hilbert space
$\ell^2_0$ of all infinite sequences of complex numbers with all but finitely
many components equal to zero. Using this fact and the continuity property
(F.\ref{cont}) of the fidelity $F$, we obtain
$$
f(\Phi,\Psi) = \inf_{\upsilon \in \fg \tp \ell^2_0; \norm{\upsilon} =
  1} F\big(\Phi_\intercal
\tp \id (\ketbra{\upsilon}{\upsilon}), \Psi_\intercal \tp \id
(\ketbra{\upsilon}{\upsilon}) \big).
$$
Using this expression in conjunction with Eq.~(\ref{eq:fg}), we get the bounds
\begin{eqnarray}
f(\Phi,\Psi) &\ge& 1-\sup_{\upsilon \in \fg \otimes \ell^2_0 :
  \| \upsilon \| = 1} D\big( \Phi_\intercal \otimes
  \operatorname{id} (| \upsilon \rangle \langle \upsilon |),\Psi_\intercal \otimes
  \operatorname{id} (| \upsilon \rangle \langle \upsilon |)\big)  \label{eq:fidbound1} \\
f^2(\Phi,\Psi) &\le& 1-\sup_{\upsilon \in \fg \otimes \ell^2_0 :
  \| \upsilon \| = 1} D^2\big(\Phi_\intercal \otimes
  \operatorname{id} (| \upsilon \rangle \langle \upsilon |) -
  \Psi_\intercal \otimes
  \operatorname{id} (| \upsilon \rangle \langle \upsilon |)\big).
\label{eq:fidbound2}
\end{eqnarray}
Now, for any completely bounded map
$\map{\Lambda}{\cB(\fh)}{\cB(\fg)}$, the image of the set $\{
\ketbra{\upsilon}{\upsilon} : \upsilon \in \fg \tp \ell^2_0,
\norm{\upsilon} = 1\}$ under the predual map $\map{\Lambda_\intercal
\tp \id}{\cB_\intercal(\fg \tp \ell^2_0)}{\cB_\intercal(\fh \tp \ell^2_0)}$ is contained in the
trace-norm closure of the linear span of $\{ \ketbra{\xi}{\xi} : \xi
\in \fh \tp \ell^2_0, \norm{\xi} = 1\}$, which is dual to the tensor
product $\cB(\fh) \tp \cK(\ell^2)$, where $\cK(\ell^2)$ is the space
of compact operators on $\ell^2$. Thus, by duality we have
\begin{eqnarray*}
&& \sup_{\upsilon \in \fg \otimes \ell^2_0 :
  \| \upsilon \| = 1} D\big( \Phi_\intercal \otimes \id (\ketbra{\upsilon}{\upsilon}),\Psi_\intercal \otimes
  \id (\ketbra{\upsilon}{\upsilon})\big) \\
&& \qquad \qquad = \frac{1}{2}\norm{(\Phi -
  \Psi) \tp \id_{\cK(\ell^2)}} \\
&& \qquad \qquad \equiv \cD(\Phi,\Psi),
\end{eqnarray*}
where $\cD(\Phi,\Psi)$ denotes the CB-norm half-distance
$(1/2)\cbnorm{\Phi - \Psi}$, and the last equality follows from the
formula $\cbnorm{\Lambda} = \norm{\Lambda \tp \id_{\cK(\ell^2)}}$ for
any completely bounded map $\Lambda$.

On the other hand, when $\dim \fg = m < \infty$, we can use the fact
\cite{Pau03} that, for any completely bounded map $\Lambda$ into $\cB(\fg)$,
$$
\cbnorm{\Lambda} = \norm{\Lambda \tp \id_{\cM_m}} =
\trnorm{\Lambda_\intercal \tp \id_{\cM_m}},
$$
where $\cM_m$ denotes the algebra of $m\times m$ complex matrices,
whence it follows that
$$
\sup_{\upsilon \in \fg \tp \fg : \| \upsilon \|} D\big(\Phi_\intercal
\tp \idty (\ketbra{\upsilon}{\upsilon}), \Psi_\intercal \tp
\idty(\ketbra{\upsilon}{\upsilon})\big) = \cD(\Phi,\Psi).
$$
In either case, we immediately derive the inequality
\begin{equation}
1-\cD(\Phi,\Psi) \le f(\Phi,\Psi) \le \sqrt{1-\cD^2(\Phi,\Psi)},
\label{eq:qofid_norms}
\end{equation}
which, when expressed in terms of the Hellinger distance
$d_H(\bullet,\bullet) := \sqrt{1-f(\bullet,\bullet)}$ as
\begin{equation}
2^{-1/2}\cD(\Phi,\Psi) \le d_H(\Phi,\Psi) \le \sqrt{\cD(\Phi,\Psi)},
\label{eq:hellcb}
\end{equation}
yields the desired property
\begin{itemize}
\item[(f.5)] the Hellinger distance $d_H(\bullet,\bullet) :=
\sqrt{1-f(\bullet,\bullet)}$ is topologically equivalent to the
CB-norm distance [cf.~Eq.~(\ref{eq:hellcb})].
\end{itemize}
This completes our survey of the basic properties of the minimax
fidelity $f$.

\section{Some examples and applications}
\label{sec:ex}

The expressions for the minimax fidelity, derived in Section~\ref{sec:evaluate} for different
kinds of quantum operations encountered in quantum information theory,
share the common feature of being set up as variational problems,
namely, as minimizations of a concave functional over a convex set. This
feature of the minimax fidelity renders the problem of computing it
amenable to robust numerical methods (see Ref.~\cite{GLN04} for
detailed discussion of numerical optimization methods for the
calculation of fidelity-like measures in quantum information
theory). However, there are instances in which the minimax fidelity
between two quantum channels can be written down in a more explicit
form. In this section we sketch some examples of such instances.

Before we proceed, we would like to remind the reader of the assumption we made in
Section~\ref{ssec:opdens}, namely that all the channels we deal with are
completely majorized by the trace in the sense of
Ref.~\cite{BeSta86}. This assumption, while allowing us to circumvent certain
technicalities involving unbounded operators, is somewhat restrictive,
as one can easily find examples of quantum channels between
infinite-dimensional algebras (e.g., unitarily or isometrically
implemented channels; see Ref.~\cite{Be02} for details) that do not satisfy this condition of
complete majorization. However, owing to the CB-continuity of the
minimax fidelity [cf.~Section~\ref{sec:properties}], we may always
regard such channels as CB-limits of sequences of channels with
finite-dimensional output algebras. Thus, given a channel
$\map{\Phi}{\cB}{\cA}$, $\cB=\cB(\fh)$, $\cA = \cB(\fg)$ with $\dim
\fg = \infty$, we consider a sequence $\{ P_n \}$ of
finite-dimensional projections such that $P_n \to \idty_\fg$ strongly,
and the corresponding sequence $\{ \Phi_n \}$ of quantum
operations $\Phi_n(B) := P_n\Phi(B)P_n$, so that $\Phi_n(B)
\to \Phi(B)$ uniformly as $n \to \infty$ for each $B \in \cB$, and each $\Phi_n$ is a channel from
$\cB$ into $P_n \cA P_n$, with $\lim_{n \to \infty}\cbnorm{\Phi -
\Phi_n} = 0$.

With this in mind, in the examples below we shall not worry about the
issue of bounded vs. unbounded operational densities.

\subsection{Unitary maps}
\label{ssec:unitary}

In the case of channels $\Theta_U,\Theta_V$ implemented by the
unitaries $\map{U,V}{\fh}{\fh}$, i.e., $\Theta_U(B) = U^\dag B U$ and $\Theta_V(B)
= V^\dag B V$, the minimax fidelity $f(\Phi,\Psi)$ is easily evaluated
using Eq.~(\ref{eq:qfidkraus}):
$$
f(\Theta_U,\Theta_V) = \inf_{\varrho \in \cS(\fg)} \abs{\varrho(W)},
$$
where we have defined $W := U^\dag V$. Let $\Sp (W)$ denote the
spectrum of $W$, which is a closed compact subset of the unit circle
$\mathbb{T}$ in the complex plane, and let $E^W(\mathrm{d}z)$ denote the
corresponding spectral measure of $W$. Then we can write
$$
f(\Theta_U,\Theta_V) = \inf_{\varrho \in \cS(\fg)} \Abs{\int_{\Sp (W)}z
M^{W,\varrho}(\mathrm{d}z)},
$$
where $M^{W,\varrho}(\mathrm{d}z)$ is the probability measure
$\varrho[E^W(\mathrm{d}z)] \equiv (E^W(\mathrm{d}z),\rho)$. Thus
\begin{equation}
f(\Theta_U,\Theta_V) = \mathrm{dist}(0,\overline{\co \Sp (W)}),
\label{eq:mm_unitary}
\end{equation}
where $\overline{\co \Sp (W)}$ denotes the closed convex hull of $\Sp (W)$, and
$\mathrm{dist}(z,S) := \inf \{ \abs{z-z'} : z' \in S \}$ for any $z \in
\C$ and $S \subset \C$. Clearly, $f(\Theta_U,\Theta_V) = 1$ if and only if
$\overline{\co \Sp W} \subset \mathbb{T}$, i.e., if and only if $W = \lambda
\idty_\fh$ with $\abs{\lambda} = 1$, which is equivalent to $\Theta_U = \Theta_V$.

When $\dim \fh < \infty$, $\Sp (W)$ is a finite subset of
$\mathbb{T}$, so that $\overline{\co \Sp (W)}$ is a polygon in the
complex plane, and Eq.~(\ref{eq:mm_unitary}) shows that $f(\Theta_U,\Theta_V)$
is simply the distance $d$ from this polygon to
the origin. On the other hand, recalling the formula \cite{AKN97}
$\cD(\Theta_U,\Theta_V) = \sqrt{1-d^2}$, we see that the upper bound
in Eq.~(\ref{eq:qofid_norms}) is saturated by the unitarily implemented channels.

\subsection{Random unitary channels}
\label{ssec:randomunitary}

Continuing with the set-up from the preceding example, let us consider
channels of the form
\begin{equation}
\Phi(B) = \sum_i p_i \Theta_{U_i}(B),\; \Psi(B) = \sum_i q_i \Theta_{U_i}(B),
\end{equation}
where $\Theta_{U_i}$ are unitarily implemented channels and $\boldsymbol{p}
\equiv \{ p_i \}$, $\boldsymbol{q} \equiv \{
q_i \}$ are probability distributions. It the follows immediately from the
strong concavity property (\ref{eq:strconcave_mm}) of the minimax
fidelity that
\begin{equation}
f(\Phi,\Psi) \ge \sum_i \sqrt{p_i q_i} \equiv F(\boldsymbol{p},\boldsymbol{q}).
\label{eq:randomunitary1}
\end{equation}

When $\dim \fh < \infty$, the inequality in (\ref{eq:randomunitary1})
becomes equality when the unitaries $U_i$ are orthogonal in the Hilbert-Schmidt sense,
$\Tr U_i^\dag U_k = \dim \fh \cdot \delta_{ik}$. On the other hand,
when $\fh$ is infinite-dimensional, this orthogonality condition does not make sense unless
we consider channels given in terms of continual Kraus decompositions,
so that the sums in Eq.~(\ref{eq:randomunitary1}) are replaced with
integrals with respect to some positive measure $\mu$, and agree to understand orthogonality in the sense of operator-valued
Schwartz distributions. As an example, consider the following.

Let $\fh = \cF$, the boson Fock space, let $a$ and $a^\dag$ be the field
annihilation and creation operators, and let $D(z) :=
\exp (z a^\dag - \bar{z} a)$, $z \in \C$, be the unitary displacement
operators obeying the Weyl relation $D(z)D(z') =
\mathrm{e}^{\mathrm{i}\Im zz'} D(z+z')$. Given a function $f \in
L^2(\C,\mathrm{d}z)$, where $\mathrm{d}z := \mathrm{d}(\Re
z)\mathrm{d}(\Im z)$, we define its {\em Weyl--Fourier transform} as $D(f) :=
{\pi}^{-1/2} \int_\C f(z)D(z) \mathrm{d}z$. Since $f$ is
square-integrable, $D(f)$ is a Hilbert-Schmidt operator, and it can be
easily shown that
$$
\Tr \big[ D(f)^\dag D(g) \big] = \int_\C \overline{f(z)}g(z)\mathrm{d}z \equiv
\langle f, g \rangle_{L^2(\C)}, \qquad \forall f,g \in L^2(\C)
$$
so that $\Tr \big[ D(z)^\dag D(z')] = \pi \delta^{(2)}(z-z')$, $z,z'
\in \C$, where $\delta^{(2)}(\lambda) := \delta(\Re \lambda)\delta(\Im
\lambda)$ is the Dirac $\delta$-function in the complex plane.

With this in mind, consider the family of
channels $\map{\Gamma^{(\mu)}}{\cB(\cF)}{\cB(\cF)}$, $\mu \in \R^+$, with
the preduals given by
$$
\Gamma^{(\mu)}_\intercal(\rho) := \frac{1}{\pi \mu} \int_\C D(z) \rho D(z)^\dag
\exp(-|z|^2/\mu) \mathrm{d}z
$$
(in quantum optics these channels model the so-called {\em Gaussian
  displacement noise} \cite{Hal94}). Then the minimax fidelity between $\Gamma^{(\mu)}$ and
$\Gamma^{(\nu)}$ is given by
\begin{equation}
f(\Gamma^{(\mu)},\Gamma^{(\nu)})=\frac{(\mu\nu)^{\frac{1}{2}}}{\tfrac{1}{2}%
(\mu+\nu)}.
\label{eq:mm_gauss}
\end{equation}
Owing to the inequality between the geometric and the arithmetic means, the
right-hand side of Eq.~(\ref{eq:mm_gauss}) is always bounded between 0
and 1, and the maximum value of 1 is attained if and only if $\mu
= \nu$, i.e., $\Gamma^{(\mu)} = \Gamma^{(\nu)}$. This, of course,
agrees with the properties of the minimax fidelity (cf.~Section~\ref{sec:properties}).

\subsection{Master equation}
\label{ssec:master}

Consider a strongly continuous semigroup of channels
$\{ \map{\Phi^{(t)}}{\cB(\fh)}{\cB(\fh)} \}_{t \in \R^+}$, with the preduals $\Phi^{(t)}_\intercal$
satisfying the Lindblad master equation \cite{Lin76}
\begin{equation}
\frac{\mathrm{d}\Phi^{(t)}_\intercal(\rho)}{\mathrm{d}t} = X \rho X^\dag -
  \frac{1}{2}(X^\dag X \rho+ \rho X^\dag X)
\end{equation}
for some $X \in \cB(\fh)$. Introducing the dilating Hilbert space $\cH = \fh \tp \fh$ with the
basis $\{ \ket{0}, \ket{1}, \ldots \}$, we can, for an infinitesimal time
$t=\varepsilon$, write the predual of
the channel $\Phi^{(\varepsilon)}$ in the Stinespring form
\begin{equation}
\Phi^{(\varepsilon)}_\intercal(\rho) = \Tr_\cH A_\varepsilon \rho A_\varepsilon^\dag,
\end{equation}
where the map $\map{A_\varepsilon}{\fh}{\fh \tp \cH}$ is
given by
\begin{equation}
A_\varepsilon \upsilon := \left( \idty_\fh - \frac{1}{2}\varepsilon
X^\dag X \right) \upsilon \tp \ket{0} + \sqrt{\varepsilon} X\upsilon
\tp \ket{1} + O(\varepsilon^2),
\end{equation}
$O(\varepsilon^2)$ indicating terms with norm bounded from above by
$M\varepsilon^2$ for some constant $M \ge 0$. Note that $A_0 \upsilon = \upsilon \tp \ket{0}$, so that $T^{(0)} = \id$.
We can then evaluate the partial trace
\begin{equation}
\operatorname{Tr}_\fh [A_{\varepsilon}\rho A_0^\dag] = \left( 1-\frac{1}%
{2}\varepsilon\langle X^\dag X\rangle_\rho \right) \ketbra{0}{0} + \sqrt
{\varepsilon}\langle X\rangle_\rho \ketbra{1}{0} + O(\varepsilon^2),
\end{equation}
where $\langle B \rangle_\rho := \Tr (B \rho)$ for $B
\in \cB(\fh)$. Then, again up to an additive term of operator norm
$O(\varepsilon^2)$,
\begin{equation}
\operatorname\Tr_\fh [A_\varepsilon \rho A_0^\dag]^\dag
\Tr_\fh[A_\varepsilon \rho A_0^\dag] \approx \left[ \left( 1-\frac{1}{2}%
\varepsilon\langle X^\dag X\rangle_\rho \right) ^2 + \varepsilon
\abs{\langle X\rangle_\rho}^2 \right] \ketbra{0}{0},
\end{equation}
which allows us to compute, up to $O(\varepsilon^2)$, the minimax
fidelity between the channel $T^{(\varepsilon)}$ after an
infinitesimal time $\varepsilon$ and the identity map. Using
Eq.~(\ref{eq:qofid_stine}), we obtain
\begin{equation}
f(T^{(\epsilon)},\id) = \inf_{\varrho\in\cS(\fg)} \trnorm{\Tr_\fh
  [A_\varepsilon \rho A_0^\dag]} \approx \sqrt{1-\varepsilon C},
\end{equation}
where
\begin{equation}
C=\inf_{\varrho \in \cS(\fg)} \big(\langle X^\dag X\rangle_\rho - \abs{\langle X\rangle_\rho}^2\big).
\end{equation}

\subsection{Impossibility of quantum bit commitment}
\label{ssec:qbc}

The statement of topological equivalence of the noncommutative
Hellinger distance and the CB-norm distance between a pair of quantum
channels, i.e., Eq.~(\ref{eq:hellcb}), is essentially the ``continuity
argument" at the heart of a proof of ``impossibility of quantum bit commitment
(QBC)" \cite{LoChauQBC97}. Quantum bit commitment is a cryptographic objective in which one party, Alice, commits a
bit to another party, Bob, in such a way that the corresponding protocol is {\em concealing}
(i.e., Bob is not able to retrieve the bit before the opening) and
{\em binding} (i.e., Alice cannot change the bit after the
commitment). The impossibility proof asserts that if the protocol is
perfectly concealing, then it is necessarily not binding, and invokes
a continuity argument for
``asymptotically" concealing protocols, stating that Alice's
probability of successful cheating approaces unity, while Bob's cheating
probability becomes close to the value
$1/2$ (pure guessing).\footnote{The reader should be aware that the 
  impossibility proof in Ref.~\cite{LoChauQBC97} is valid for a
restricted class of protocols, i.e., those that are non-aborting and have a single commitment step. For wider
  classes of protocols, it is still a matter of debate whether a secure
QBC protocol exists \cite{Yuen}.}  In this example we derive the
continuity argument from the expression of Alice's and Bob's
respective cheating
probabilities as a consequence of the topological
equivalence between the Hellinger distance and the CB-norm distance in
Eq. (\ref{eq:hellcb}).

From the point of view of Bob, Alice's action of committing the bit is
equivalent to a channel $\Phi_{\boldsymbol{A}^{(b)}}$ on an algebra
$\cB(\fh)$, $\dim \fh < \infty$,  for each
value of the committed bit $b=0,1$, where $\boldsymbol{A}^{(b)} \equiv
\{ A_j^{(b)}\}^k_{j=1}$ is a collection of operators satisfying the Kraus
condition $\sum^k_{j=1} {A_j^{(b)\dag}} A_j^{(b)} = \idty$, and
$\Phi_{\boldsymbol{A}^{(b)}}$ denotes the channel induced by this
Kraus decomposition. At the opening, Alice informs Bob about which
element of the Kraus decomposition $\boldsymbol{A}^{(b)}$ she actually
used in
the commitment. However, prior to unveiling the label $j$, Alice can
perform an {\em EPR attack} with the purpose of changing the Kraus decomposition to another
equivalent decomposition $\boldsymbol{A}^{(b)}(V) \equiv \{A_j^{(b)}(V)\}$, where
$A_j^{(b)}(V):= \sum_\ell A_l^{(b)} V_{j\ell}$ for some $V\in {\sf U}(\C^k)$.  The EPR attack
is achieved by Alice via the unitary transformation $V$ on an
ancillary $k$-dimensional space $\cH$. The conditional probability that
Alice can cheat successfully by convincing Bob that she has committed, say,
$b=1$, while having successfully committed $b=0$ instead, is given by 
\begin{equation}
P_c^A(V,\upsilon)=\sum_j 
\frac{\abs{\bra{\upsilon} A_j^{(0)}{}^\dag(V)A_j^{(1)}\otimes \idty_\cH
 \ket{\upsilon}}^2}{\norm{(A_j^{(1)}\otimes \idty_\cH)\upsilon}^2},
\label{cheatA}
\end{equation}
where $\norm{(A_j^{(0)}\otimes \idty_\cH)\upsilon}^2$ is the
probability that the $j$th Kraus element is unveiled.  Which $V$
should Alice use? Without any knowledge of $\ket{\upsilon}$, the best
she can do is to adopt a conservative strategy of choosing the $V$
that will maximize her cheating probability in the worst-case scenario,
namely for the anonymous state $\ket{\upsilon}$ chosen by Bob to
minimize $P_c^A(V,\upsilon)$. This is the {\em minimax} choice of $V$,
corresponding to the cheating probability 
\begin{equation}
\bar{P}_c^A := \sup_{V\in {\sf U}(\C^k)}\inf_{\upsilon\in \fh\otimes\cH; \norm{\upsilon}=1}
    P_c^A(V,\upsilon),
\label{PAc}
\end{equation}
On the other hand, for equiprobable bit values $b \in \{0,1\}$ Bob's optimal probability of
cheating is given by the probability of error in discriminating
between the
corresponding output states, more precisely
\begin{equation}
\bar{P}_c^B = \frac{1}{2} + \frac{1}{4} \sup_{\upsilon \in \fh
\otimes \cH; \norm{\upsilon}=1}
\trnorm
{
\rho^\upsilon_{\boldsymbol{A}^{(0)}} - 
\rho^\upsilon_{\boldsymbol{A}^{(1)}}}
= \frac{1}{2}
\left[1+ \cD(\Phi_{\boldsymbol{A}^{(0)}},\Phi_{\boldsymbol{A}^{(1)}})\right],
\label{PBc}
\end{equation}
where we have defined $\rho^{\upsilon}_{\boldsymbol{A}} := \Phi_{\boldsymbol{A}}
\otimes\id (\ketbra{\upsilon}{\upsilon})$. Using Jensen's
inequality, we can bound Alice's cheating probability $P_c^A(V,\upsilon)$
from  below as
\begin{equation}
P_c^A(V,\upsilon)\ge\left|\sum_j 
|\langle\upsilon|A_j^{(0)}(V){}^\dag A_j^{(1)}
\otimes \idty_\cH|\upsilon\rangle|\right|^2.
\label{boundPAc}
\end{equation}
Note that the value of the max-min in Eq.~(\ref{PAc}) will not change
if we perform the maximization over the
closed convex hull of ${\sf U}(\C^k)$, i.e., the set ${\sf K}(\C^k)$ of
all linear contractions on $\C^k$, and the minimization over the closed convex
hull of the pure states on $\fh \tp \cH$, i.e., the set $\cS(\fh \tp
\cH$) of states on $\cB(\fh \tp \cH)$, thus completing the domain of the max-min to the product
${\sf K}(\C^k) \times \cS(\fh \tp \cH)$ of compact convex sets. Now, the
functional
\begin{equation}
F(V, \rho) :=
\sum_j\Re \Tr\{\rho [A_j^{(0)}(V){}^\dag A_j^{(1)} \otimes \idty_\cH]\}
\end{equation}
is affine in both $V \in {\sf K}(\C^k)$ and $\rho \in \cS(\fh\tp \cH)$,
so that we can use standard minimax arguments \cite{Lue69} to
justify the interchange of extrema in Eq.~(\ref{PAc}), and then apply Lemma~\ref{lm:homf} to obtain
\begin{eqnarray}
&& \sup_{V\in {\sf U}(\C^k)}\inf_{\upsilon\in\fh\otimes\cH}
|F(V,|\upsilon\rangle\langle\upsilon|)| =
\sup_{V\in {\sf K}(\C^k)}\inf_{\varrho\in\mathcal{S}(\fh\otimes\cH)}
|F(V,\rho)|\\
&& \qquad \qquad = \inf_{\varrho\in\mathcal{S}(\fh\otimes\cH)}\sup_{V
  \in {\sf K}(\C^k)}
|F(V,\rho)|\\
&& \qquad \qquad =
\inf_{\upsilon\in\fh\otimes\cH; \norm{\upsilon}=1}\sup_{V\in {\sf U}(k)}
|F(V,|\upsilon\rangle\langle\upsilon|)|.
\label{minmaxf}
\end{eqnarray}
Now, since a monotone function does not affect the saddle point, we can use Eqs.~(\ref{eq:minimax2a}), (\ref{PAc}), (\ref{boundPAc}), and
(\ref{minmaxf}) to obtain
$$
\bar{P}_c^A \ge f^2(\Phi_{\boldsymbol{A}^{(0)}},\Phi_{\boldsymbol{A}^{(1)}}).
$$
Using Eq.~(\ref{eq:hellcb}) and then Eq.~(\ref{PBc}), we finally
obtain the chain of estimates
$$
\bar{P}_c^A \ge
f^2(\Phi_{\boldsymbol{A}^{(0)}},\Phi_{\boldsymbol{A}^{(1)}}) \ge
[1-\cD(\Phi_{\boldsymbol{A}^{(0)}},\Phi_{\boldsymbol{A}^{(1)}})]^2 
\ge [1-2(\bar{P}^B_c-1/2)]^2, 
$$
whence it follows that, for ``asymptotically'' concealing protocols,
i.e., those for which $\bar{P}_c^B \to \frac{1}{2}$, Alice's probability of cheating will
approach unity, and the protocol will not be binding. 

\section*{Acknowledgments}
\noindent
This work has been sponsored by the Multiple Universities Research
Initiative (MURI) program administered by the
U.S. Army Research Office under Grant
No. DAAD19-00-1-0177. V.P.B. ackowledges support from EC under the program ATESIT (Contract No.
IST-2000-29681). G.M.D. also acknowledges support by EC and Ministero Italiano dell'Universit\`a e della  
Ricerca (MIUR) through the cosponsored ATESIT
project IST-2000-29681 and Cofinanziamento 2002. M.R. acknowledges
kind hospitality of the Quantum  
Information Theory Group
at Universit\`a di Pavia, and support from MIUR under
Cofinanziamento 2002 and from the European Science Foundation. The
authors would like to thank the referee for several suggestions, which
resulted in improved presentation.

\end{document}